\documentclass[a4paper,10pt]{article}%''draft'' is added to allow compilation sans figures (WA)
\usepackage[hmargin=2.42cm, vmargin=2.42cm]{geometry} 
\usepackage{amssymb,amsmath,amsthm,enumerate}
\usepackage{array}
\usepackage{wrapfig}
\usepackage{fancyhdr}
\usepackage{fancyvrb}
\usepackage{color,array,rotating}
\usepackage{graphicx}
\usepackage{slashbox} % I don't have this package [Wojtek]
\usepackage{cite}
\usepackage{setspace}
\usepackage{multirow}

\usepackage{booktabs}

\usepackage{color}
\definecolor{vincent}{rgb}{0.5,0.5,0.5}
\definecolor{motta}{rgb}{0,0,0}
\newcommand{\red}[1]{\textcolor{black}{#1}}
\newcommand\compare[5]{\ifthenelse{\lengthtest{#1pt #2 #3pt}}{#4}{#5}}
\renewcommand\Re{{\rm Re}}
\newcommand\We{{\rm We}}
\newcommand\Refi{{$1.37 \, 10^4$}}
\newcommand\Wefi{{$41.6$}}
\newcommand\Refii{{$4.33 \, 10^5$}}
\newcommand\Wefii{{416}}
\newcommand\Order{{\cal O}}
\begin{document}
\begin{centering}
\centerline{\textbf{\Large{A phase inversion benchmark
}}} \centerline{\textbf{\Large{for multiscale multiphase flows}}}
%\vspace{0.6cm} \noindent
%\centerline{\textbf{\Large{Part I: numerical modeling}}}        % Enter your title between curly brace
\vspace{0.5cm} \noindent

J.-L. Estivalezes$^{2,4}$,  W. Aniszewski$^3$, F. Auguste$^{4}$, Y. Ling$^{5,6}$,\\  L. Osmar$^7$,  J.-P. Caltagirone$^7$, L. Chirco$^5$, A. Pedrono$^4$, S. Popinet$^5$, \\A. Berlemont$^3$, J. Magnaudet$^4$, T. M\'enard$^3$, S. Vincent$^9$, S. Zaleski$^{1,5,10}$\\

\noindent \vspace*{10pt} \noindent

$^1$ \textit{stephane.zaleski@sorbonne-universite.fr (corresponding author)}\\

\noindent \vspace*{10pt} \noindent

$^2$ ONERA, The French Aerospace Lab, F-31055 Toulouse, France, \\
$^3$ Universit\'e de Rouen and CNRS, Complexe de Recherche Interprofessionnel en A\'erothermochimie (CORIA) UMR 6614,  F-76801 Saint-Etienne-du-Rouvray Cedex, France, \\
$^4$ Institut de M\'ecanique des Fluides de Toulouse (IMFT), Universit\'e de Toulouse, CNRS, Toulouse, France,\\
$^5$ Sorbonne Universit\'e and CNRS, Institut Jean Le Rond d'Alembert UMR 7190, F-75005 Paris, France, \\
$^6$ Baylor University, Department of Mechanical Engineering, Waco, TX 76798, USA,\\
$^7$ Bordeaux INP, University of Bordeaux, CNRS, Arts et M\'etiers Institute of Technology, INRAE, Institut de M\'ecanique et Ing\'enierie (I2M) UMR 5295, F-33400 Talence, France,\\
%$^8$ CERFACS, Centre Europ\'een de Recherche et de Formation Avanc\'ee en Calcul Scientifique, F-31057 Toulouse, France, \\
$^9$ Universit\'e Paris-Est Marne-La-Vall\'ee and CNRS, Laboratoire Mod\'elisation et Simulation Multi Echelle (MSME), UMR 8208, F-77454, Marne-La-Vall\'ee, France, \\
$^{10}$ Institut Universitaire de France, Paris, France. \\

\vspace*{0.5cm} \noindent
\end{centering}
\begin{abstract}
  A series of benchmarks based on the physical situation of ``phase inversion" between two {immiscible} %formerly incompressible
  liquids is presented. These benchmarks aim at progressing towards the direct numerical simulation of two-phase flows. Several CFD codes developed in French laboratories and using either Volume-of-Fluid or Level-Set interface tracking methods are {used} to provide physical solutions of the benchmarks, convergence studies and code comparisons. Two typical configurations are retained, with integral scale Reynolds numbers of \Refi \mbox{} and \Refii, respectively. The physics of the problem are probed through macroscopic quantities such as potential and kinetic energies, or enstrophy. In addition, scaling laws for the temporal decay of the kinetic energy are derived to check the physical relevance of the simulations. Finally the droplet size distribution is probed. Additional test problems are also reported to estimate the influence of viscous effects in the vicinity of the interface.
\end{abstract}


\newpage

\section{Introduction}

The physical validation \textcolor{black}{or verification} of numerical codes devoted to multiphase flows and immiscible fluids (by immiscible we mean that interfaces remain sharp) is a major concern of modern CFD. A lot of work was carried out since the 1990s to validate interface-tracking algorithms in cases where the velocity field is given analytically \cite{Rider,Rudman}. Comparisons have also been achieved with available experiments or predictions of linear stability theories \cite{WebTestCase}. However, in these situations, the topology of the interface is simple and the flow problem involves {a single scale and a single physical effect} such as the continuity of stresses across the interface with the two-phase Poiseuille flow \cite{Benkenida}, the Laplace law for a spherical or circular drop\cite{Popinet}, the oscillation modes of a free surface in the linear regime\cite{Prosperetti3,Popinet}, the head-on coalescence of two drops \cite{Ashgriz,Tanguy},
%the dam break problem \cite{Martin}
or the break-up of a low-velocity round jet due to the Rayleigh-Plateau instability \cite{Delteil} to mention just a few.

\begin{figure}[ht!]
\begin{center}
  \includegraphics[width=6cm]{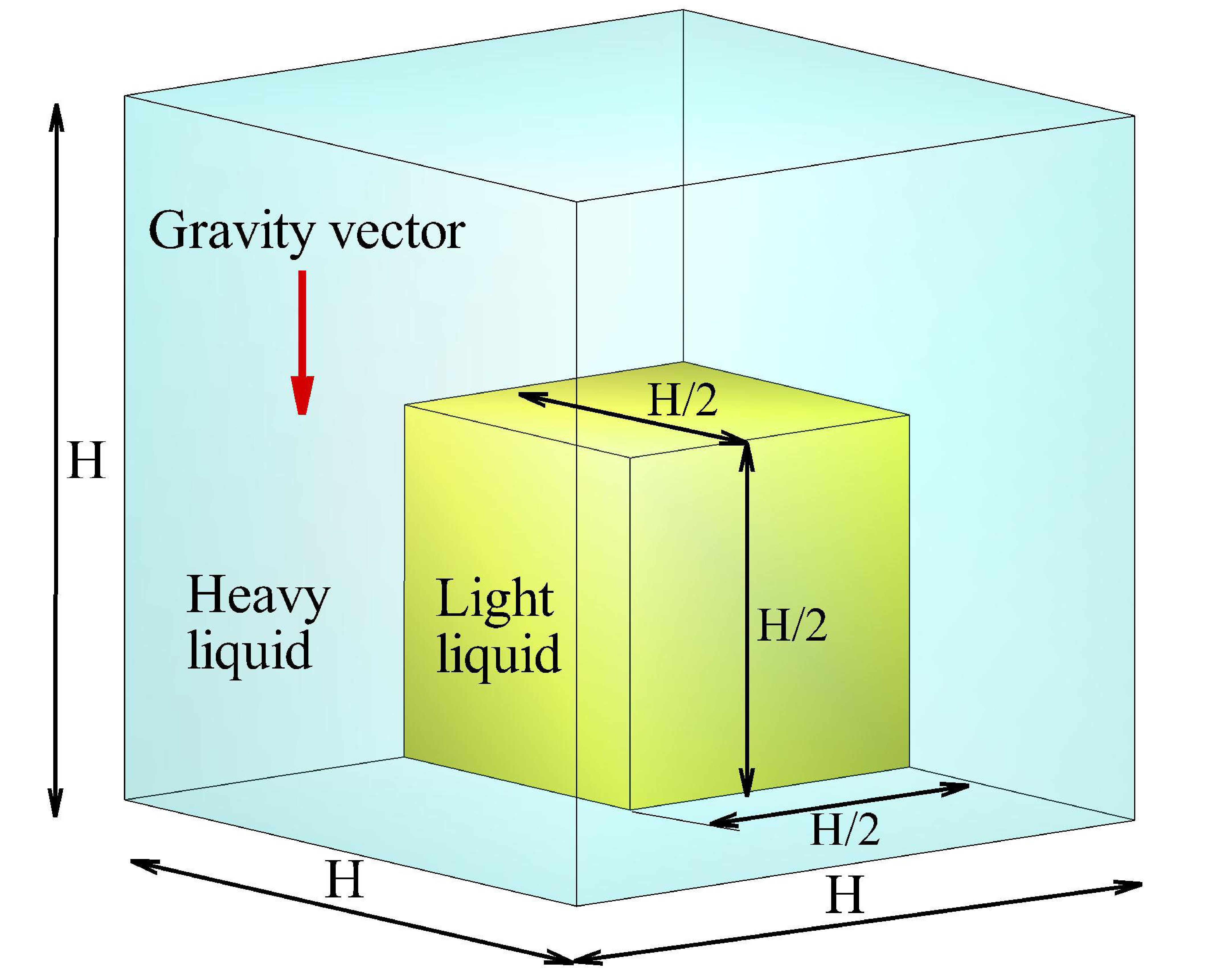}
  \end{center}
\caption {Initial setup of the phase inversion problem in a closed box. The light liquid or oil is fluid 1 and the more dense, less viscous liquid is fluid 2.}
\label{fig2.1}
\end{figure}

None of these problems puts into play the interaction between fluctuating interfaces and a turbulent flow, which is a key point in many realistic cases involving a two-phase dynamics. Indeed, such multiphase interactions occur in many environmental and industrial applications. Let us just mention the atomization of a liquid jet in an engine\cite{lasheras00,ling2019two,eggers08}, phase separators in chemical engineering processes, the ladle-based elaboration of steel, the boiling crisis in nuclear reactors\cite{gilman2017self}, complex bubbly flows including the flow past Taylor bubbles in pipes\cite{lu2018direct}, droplet or asteroid impacts\cite{josserand2016drop}, wave impact on structures or the ubiquitous phenomenon of wave breaking \cite{Deike:2016ir,mostert2021high,mostert2020inertial}. In most multiphase flows of environmental or engineering relevance, the multiscale character of the flow is a key issue to be handled by computational approaches. The interfacial length scales are associated on the one hand to large interface structures such as jets, films or large drops. On the other hand, small interfacial scales also exist, corresponding to a small-scale dispersed phase, {\it i.e.} small bubbles or droplets. These two widely distinct families of interfacial scales interact in a nonlinear way through ligament break-up or drop/bubble coalescence and the unsteady or turbulent nature of the carrier fluid motion plays a crucial role in these interactions. The definition of a reference benchmark for these multiscale sharp interface problems appears to be an important issue. \\

\begin{figure}[ht!]
\begin{center}
  \includegraphics[width=6cm]{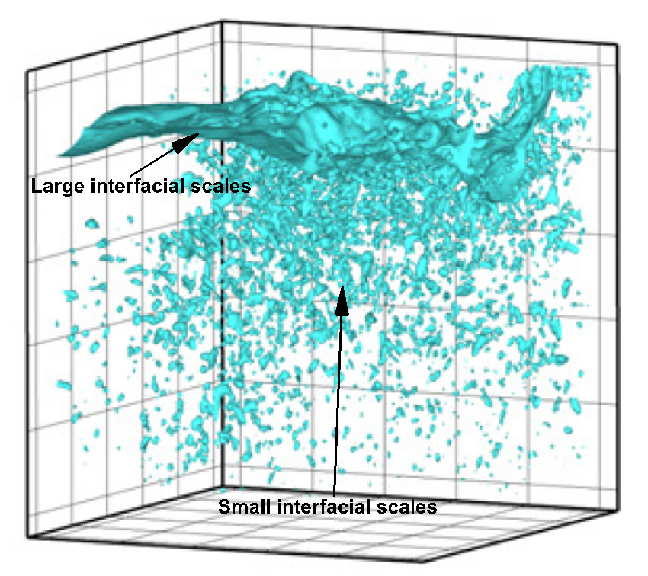}
  \end{center}
\caption {Example of simulated phase inversion in a closed box \cite{Vincent4}  illustrating the multiscale character of the interfacial flow. (The case is similar to our case 2 but with smaller surface tension.)}
\label{fig1.1}
\end{figure}

Homogeneous Isotropic Turbulence (HIT) often serves as a reference
problem in the \textcolor{black}{verification} of single-phase DNS
codes. In analogy, one would like to define a reference problem for
the \textcolor{black}{verification} of CFD multiphase flow codes devoted
to immiscible fluids. Our first idea was to try to
  define the reference problem on the basis of a well-documented
  experiment, such as atomization of liquid sheets or jets. However,
  the quality of the modeling and simulation of atomization is very
  sensitive to {inflow} conditions, in particular upstream
  turbulence, boundary layers or inlet pump oscillations
  \cite{ling17,Matas:2018fw}. Although numerous experimental studies can
  be found in the literature \cite{Martin,Lance,Chanson} they are plagued by two difficulties that make them
  unsuitable as references cases: either they have too simple flow
  structures (as for dam break on a dry bottom) or they are not
  sufficiently well documented or instrumented to provide all the
  information required for properly defining boundary and initial conditions
  for simulations. Moreover, many experiments lack instantaneous
  pressure, velocity and interface measurements (dam break on a wet
  bottom \cite{Kleefsman}).
  %
  %  --- we have decided ---
  %
  As a consequence, we decided to define
  a synthetic reference case unrelated to a specific experiment.
  It is the
{\em phase inversion} of an initial cube of light
fluid inside a larger cube of a heavier fluid, both contained in a closed box (Figure \ref{fig2.1}).
(Although an experiment in this configuration could in principle be
  considered, the setup resembles that of an oil and water phase separator or settling pond, or of a model Rayleigh-Taylor experiment, and with a $90^\circ$-change in the gravity orientation, that of the dam-break problem \cite{Martin}.)
  In this reference case, initial and boundary conditions
  are simple and well defined. {We also simplify the problem somewhat by using
  a liquid-liquid density ratio ($\rho_1/\rho_2=9/10$). There are two reasons
  for this simplification: 1) with the air-water density ratio, the air inertia is
  not large enough to defeat the liquid inertia and gravity (this is why the dam-break problem
  is not producing much atomization); and 2) the numerical issues are considerably simplified
  with moderate density ratios. (We note that a study of atomization with
  liquid-liquid density ratios was published recently \cite{constante2021direct}. The present study
  aims at the same kind of atomization but ``in a box''.)}
  We wish to have complex
  flow features such as turbulence, interface rupture, coalescence,
  and multiple scales. Typically, the higher the Reynolds number $\Re$ and Weber number $\We$
  the more interesting and complex will the flow be. 
  Moreover, we note that in single-phase turbulence there is a 
 large-Reynolds-number state where the range of scales is very wide and 
  the statistical features of the flow on large and
  intermediate scales become relatively independent of $\Re$, a kind of {\em universality}.
  Is there such an equivalent universal state at large $\Re$ and $\We$ in multiphase turbulence? This is not certain but there
  is clearly a benefit to reach an analogous state there. 
  Such a fully-developed multiphase turbulence would be the relevant
flow state with respect to common natural
sciences or engineering situations.
An example of the phase inversion flow at very high  $\Re$ and $\We$
in which a broad range of interfacial scales coexist, multiple break-up and coalescence
events take place, and large-scale sloshing motions are observed together with small
dispersed droplets interacting with larger drops, is illustrated
in Figure \ref{fig1.1} (taken from \cite{Vincent4,Vincent64}).

{However, with very high Re and We numbers, 
 viscosity is considerably smaller than what is required to
 obtain  fully-resolved simulations in multiphase as well as in
 single-phase flow and one cannot hope to perform truly a
   Direct Numerical Simulation (that is, a simulation with all length scales
   resolved and quantities such as dissipation and enstrophy converged). There is thus a dilemma or
   a balance to strike between the interest for ``fully developed'' multiphase turbulence
   and the desire to have a fully converged    Direct Numerical Simulation  for reference purposes.
   We chose the parameters given below in Table \ref{tab:nondim}. In retrospect, once simulations
   are performed at great costs, it appeared clear (especially from the investigation of enstrophy convergence and 
   particle size Probability Density  Functions (PDF)  as we shall see) that the current work inclines towards the high-Re 
   side of the dilemma. We define two simulation cases in the phase inversion setup.
   Our simulations, especially in the second case (which are performed without explicit turbulent
   or subgrid droplet modeling), are finely-resolved implicit Large Eddy Simulations or Detailed Numerical Simulations.} (Note
   that other combinations of parameters were used  in \cite{Labourasse,Vincent18,Vincent4} and 
   \cite{Vincent64}.)
   We also note that the second case presented in this work is a good illustration of an atomization process
   although at higher Re and We than previous works on atomization that got closer to
 the ``converged'' side of the dilemma \cite{Menard,Fuster,ling17,ling2019two}. In a way, one may see the second case of the current problem
 as the study of ``atomization in a box''. An unexpected discovery of this study is that the
 non-convergence of certain flow characteristics is actually not only due to large Re and We numbers but also to the
 dependence of thin sheet breakup on the grid size, as shown in the analysis of droplets sizes in case 1 below.
 
Within the literature on high Re turbulent multiphase flow simulation one may distinguish two categories of works:
1) on diffusive interfaces and
2) on interfaces remaining sharp over time (immiscible
fluids). The first class of works has investigated for example the
interaction between Rayleigh-Taylor (RT) instabilities and turbulence
\cite{Cook, Ramaprabhu, Boffetta, Abarzhi} and a
benchmark has been published in 2004 by Dimonte and co-workers
\cite{Dimonte}. This reference work involves physical phenomena
similar to those occurring in the phase inversion configuration
proposed here. Much attention has been paid to characterizing, with
various models and numerical methods, the effects of initial
perturbations of the interface, volume fraction profiles, self-similar
bubble dynamics or mixing behavior. However, the major difference between reference \cite{Dimonte} and the current work is the immiscible nature of  the fluids and the discontinuous character
of the interfaces (of course from a macroscopic point of view). It has been clearly
demonstrated \cite{Abarzhi} that the diffuse character of the fluids
has a major effect on the global dynamics of the turbulent two-phase
flow in terms of spikes position for example, and that this effect
increases over time. In diffuse modeling of RT flows,
simulations mainly involve gas stratifications whereas our focus here is
on immiscible fluids (with at least one liquid phase in the considered problem).
Our
benchmark thus belongs to the second category of problems.
The phase inversion configuration has previously been investigated
 \cite{Vincent18} without the objective of analyzing the physical
 aspects of the multiscale character of the flow, nor comparing
 various interface tracking methods, but with the purpose of
 obtaining {\it a priori} estimates of subgrid terms that appear
 and must be modeled when LES is used in the framework of two-phase
 flows of immiscible fluids \cite{Wojtek}.
We note that for sharp interfaces and complex flows,  RT
simulations also exist \cite{Tryggvason2,
  Aniszewski201452}, and intermediate modeling standing between diffuse and sharp
interface approaches have also been proposed for simulating RT
instabilities \cite{Jacqmin}.
The major concern was the model and
numerical methods and there was no consideration of the turbulence
characteristics of the flows. To our knowledge, neither benchmark nor
detailed simulation works have been published about RT interaction
with turbulence in the framework of sharp liquid-liquid interfaces.

The study reported here has two objectives.
The first is to analyze convergence of the simulations
with grid refinement, thus producing a reference set of results.
Since no  experimental measurements
seem to exist for the phase inversion problem considered
here, this first objective is a necessary step.
The second objective is to compare the results of 
five distinct in-house
DNS multiphase flow codes to the reference results. The results should
highlight the
capabilities but also the limitations of up-to-date numerical methods
and interface models to tackle multiscale, multiphase, flow problems.
The outline of the paper is as follows: the phase inversion setup is defined in
the following Section 2. The codes and the numerical methods are described in Section 3.
The reference
simulations and results are in Section 4 while the results of all the codes are in Section 5, followed
by concluding remarks in Section 6. 

\section{Definition of the phase inversion benchmark}

We define the benchmark in this Section, following the motivations given in the previous section. 

\subsection{Geometry, initial and boundary conditions}

As described in Figure \ref{fig2.1}, an initial cubic blob of light liquid, referred to as fluid 1, is placed in the bottom part of a cubic box filled with a heavier liquid, referred to as fluid 2. The size of the box is $(H, H, H)$, while the size of the blob of light fluid is $(H/2, H/2, H/2)$. All the outer walls are considered as free-slip impermeable walls, so that the normal velocity is zero and the tangential components obey a symmetry, \textit{i.e.} shear-free, condition. A $\pi/2$ static contact angle is \red{implicitly prescribed on the walls through the boundary condition imposed on the volume fraction (in the VOF approach) or the level-set approach.}
\red{This is implemented  by enforcing a null Neumann (symmetry) boundary condition 
${\bf n} \cdot \nabla \phi = 0$, where $\phi$ is either the volume fraction or the level-set function.}
The gravitational acceleration is chosen as ${\bf g} = (0, -9.81, 0)$.
%The interest of this configuration is to be defined by simple initial and final flow solutions. In particular, fluid 1 must be at rest in the top part of the cavity while fluid 2 have to be located in the bottom part of the box at the end. The resulting interface must be flat and perpendicular to the gravity direction for later times.
Besides the simplicity of the initial and final configurations of the
phase inversion problem, another of its advantages is that it provides
the possibility to observe multiple coalescence and break-up events,
although that introduces an additional complexity since in a
simulation, coalescence is conditioned by numerical rather than
physical properties. %The main disadvantage of this benchmark case is that no experimental data are vailable.
%The main motivation for investigating a numerical benchmark of CFD codes is to have some possible comparison of simulations in order to provide reference results and a thorough numerical study of multi-phase flows.

\subsection{Physical parameters of the test cases}

Two test cases with different characteristics of the fluids are considered. The fluid properties
may be expressed in terms of \Re \mbox{} and \We. In
order to define these numbers, a velocity scale is required. We
obtain it from energy considerations.  In all of what follows, we use the characteristic or color function $C$ that is defined  \textcolor{black}{with respect to fluid 1}, so that $C=1$ in fluid 1 and $C=0$ in fluid 2. The potential energy of phase $n$ ($n=1,2$) is
\begin{equation}
  E_{p,n}=\int_\Omega C_n \rho_{n} gy \, {\rm d}V \,,
\end{equation}
where $C_n$ is the characteristic function of phase $n$, \textit{i.e.} $C_1 = C$, $C_2=1-C$, 
and $g= \| {\bf g} \|$.
The kinetic energy of phase $n$ is
\begin{equation}
  E_{k, n}=\frac{1}{2} \int_\Omega C_n \rho_{n} {\bf u}^2 \,{\rm d}V \,,
\end{equation}
where $ {\bf u}$ stands for the velocity field. 
\textcolor{black}{ Due to the simple topology of the interface in the initial and
final stages, the potential energies are trivially found.
At $t=0$,  $E^i_{p, 1}=\rho_1 g{H^4}/{32}$ and $E^i_{p, 2}=15 \rho_2 g {H^4}/{32}$
while for $t \rightarrow \infty$,
$E^f_{p,1}=15 \rho_1 g {H^4}/{128}$ and $E^f_{p,2}=49 \rho_2 g {H^4}/{128}$.
An upper bound on the kinetic energy is}
\begin{equation}
  E_{k,1} + E_{k,2} \le E^i_{p,1} + E^i_{p,2}  - E^f_{p,1} - E^f_{p,2} = \frac{11}{128} (\rho_2-\rho_1){gH^4}\,,
\end{equation}
\textcolor{black}{and an upper bound on the velocity is $|| {\bf u }||_2 < (2 E_k/\rho_1)^{1/2}$. (Note that the initial
  surface energy $3\sigma H^2/4$ provides a small contribution to the total energy; however we neglect surface energy in this
  reasoning for simplicity.) Identifying
the velocity with the upper bound, we obtain the velocity scale}
$$
U_K = \frac{\sqrt{11}}{8} \sqrt{ \frac{\rho_2-\rho_1}{\rho_1}gH} \,.
$$

\textcolor{black}{The dimensionless numbers are then defined based on $U_K$ and properties of fluid 2, namely}
\begin{eqnarray}
\Re=\frac{\rho_2 H U_K}{\mu_2}\,, \\
\We=\frac{\rho_2 H U_K^2}{\sigma}\,.
\end{eqnarray}
{As is customary in atomization problems, the Reynolds and Weber numbers may be combined in a way that eliminates the velocity scale, yielding the Laplace number La=$\sigma \rho_2 H/\mu_2^2=\Re^2/\We$.}

{To allow comparison with previous work on this setup,
we may also adopt the same velocity scale as in previous studies by some of us \cite{Vincent4, Vincent64, Wojtek}, namely ${U_g} = {{(\rho_2 - \rho_1)}/{\rho_1} \sqrt{{gH}/{2}}}$, and a characteristic time scale $t_c = H/(2U_g)$.}

%  For the sake of simplicity, the physical characteristics of fluid 2 are kept frozen and only those of fluid 1 are changed. ???? sz ?????

{The viscosities and densities considered here correspond approximately to a light oil or hydrocarbon
  (fluid 1) and water (fluid 2).  These configurations are interesting
  in so far as they provide interface and turbulent characteristics similar to
  those of jet atomization with simpler and more easily
  controlled initial conditions, and are as a result more convenient for
  numerical simulations. These liquid-liquid two-phase flows make it
  possible to select easily a Reynolds number range smaller, or
  even equivalent, to those encountered in real atomization
  processes while exhibiting equivalent dynamics. The characteristics
of the fluids and related dimensionless numbers are given in Tables
\ref{tab:dim}, \ref{tab:dimcases} and \ref{tab:nondim}. Case 1
corresponds to a configuration with moderate fragmentation.
The box used in case 2 has a larger size and thus yields a larger Reynolds number (by a factor of
30 compared to case 1).
The Weber number is also higher in case 2, (by a factor of 10
compared to case 1) but is increased less than the
Reynolds number since surface tension was also artificially
increased. This was done specifically to avoid having an exceedingly large
Weber number which would lead to excessive fragmentation of the interface, making it even harder than it is now to capture most of the small-scale structures and droplets.}

{The small scales of the flow may be characterized by the Kolmogorov scale
$\eta_K = \Re^{-3/4} H$ and the Hinze scale $\eta_H = \We^{-3/5} H$. Both scales are
given in Table \ref{tab:nondim}. While the Kolmogorov scale is very small, and prevents any attempt at DNS 
with the grids used in the paper (up to 2048 grid points in each direction), the Hinze scale is larger, and as
$\eta_H/H > 1/100$, it should be easily possible to resolve droplets at the Hinze scale. We note, however, that
droplet sizes much smaller than the Hinze scale have been observed in numerical simulations of atomization \cite{ling2019two}
and that one can estimate the diameter of the Taylor-Culick rims involved in the atomization process to be $d_{T} = \We^{-1} H \ll \eta_H$ (see \cite{marcotte2019density} for a detailed derivation of $d_T$ and its significance). Notice that for both reference simulations below, the cell size $\Delta x$ is such that 
$\Delta x < d_T/4$.}

\begin{table}[ht]
\begin{center}
\begin{tabular}{ccccc}
\hline
\hline
 $\rho_1$  & $\rho_2$ & $\mu_1$ & $\mu_2$ & $g$ \\
\hline
 ($kg.m^{-3}$)  & ($kg.m^{-3}$) & ($Pa.s$) & ($Pa.s$) & $(m.s^{-2})$ \\
\hline
  900 & $1000$ & 0.1 & $0.001$  & 9.81 \\
\hline
\hline
\end{tabular}
\end{center}

\caption{Fluid and physical properties common to both cases.\label{tab:dim}}
\end{table}

\begin{table}[ht]
\begin{center}
\begin{tabular}{ccc}
\hline
\hline
 Case  & $\sigma$ & $H$ \\
\hline
& ($kg.s^{-2}$)  & $(m)$ \\
\hline
 1 & 0.045 & 0.1 \\
 2 & 0.45  & 1\\
 \hline
 \hline
\end{tabular}
\end{center}

\caption{Fluid and physical properties that depend on the considered case\label{tab:dimcases}.}
\end{table}

\begin{table}[ht]
\begin{center}
\begin{tabular}{ccccccc}
\hline
\hline
 Case  & $\Re$ & $\We$ & La=$\Re^2/\We$ & $\eta_K/H$ & $\eta_H/H$ & $d_T/H$ \\
\hline
 1 & \Refi & \Wefi &  $4.5 \, 10^6$ & $7.9 \, 10^{-4}$ & 0.107 & 0.02 \\
 2 & \Refii  & \Wefii &  $4.5 \, 10^8$  & $5.9 \, 10^{-5}$ & 0.03 & 0.002 \\
\hline\hline
\end{tabular}
\end{center}

\caption{{ Dimensionless numbers for the considered cases. $\eta_K$ is the Kolmogorov scale at which inertia balances viscous stresses in a cascade, $\eta_H$ is the Hinze scale at which inertia balances capillary effects in a cascade,
and $d_T$ is the diameter of a Taylor-Culick rim. \label{tab:nondim}}}
\end{table}

%& $\sigma$ & $H$ \\

\subsection{Quantities of interest}

\label{Macro}

Several  quantities characterizing the evolution of the flow field are of primary interest. Among these physically relevant quantities we select the following.

\begin{itemize}

\item The potential energy which helps to measure the advancement of the relaxation from a high energy state with the light fluid located below the heavy fluid
  to the equilibrium state with the light fluid on top,  and
provides a characterization of the density stratification. 

\item The kinetic energy which characterizes the agitation of the system and allows to monitor the amount of dissipation.

\item The total mechanical energy minus the surface energy which is the sum of the previous two. As the Weber numbers are large, the surface energy is negligible and the
  above sum characterizes well the dissipation of energy in the system. It can also be used to characterize how far the system is from the possible universal high Re and We state discussed above. 

\item The time evolution of the volume integral of the enstrophy in both fluids. This quantity is defined as $E_{r, n}= \frac{1}{2} \int_\Omega C_n\omega^2 \,{\rm d}V$, with $\omega=\nabla \times {\bf u}$ denoting the vorticity.

\end{itemize}

\section{Model and numerical methods}

As is now well established, incompressible two-phase flows involving fluid-fluid interfaces and Newtonian fluids can be modeled by a single set of incompressible Navier-Stokes
equations with \textcolor{black}{phase-specific densities and viscosities} and
capillary forces, together with the transport equation of the phase function $C$. The resulting model takes implicitly into account
the mass and momentum jump relations at the interface \cite{Delhaye,Scardovelli}, whereas the continuity of the fluid-fluid and fluid-solid interfaces is
taken into account by the $C$ equation. The entire set of equations reads
\begin{eqnarray}
\nabla \cdot {\bf u}&=&0 \label{eq1}\,, \\
\rho\left[\frac{\partial {\bf u}}{\partial t}+ ({\bf u} \cdot{\nabla})
{\bf u}\right]&=& -\nabla p + \rho {\bf g} +\nabla \cdot \left[\mu\left(\nabla
{\bf u}+(\nabla {\bf u})^\text{T}\right)\right] + {\bf F}_{st} \label{eq2}\,, \\
\frac{\partial C}{\partial t}+ {\bf u} \cdot \nabla C &=& 0\,,
\label{eq3}
\end{eqnarray}
where $p$ is the pressure, ${\bf F}_{st}$ is the interfacial force per unit volume and $\rho$ and $\mu$ are the local density and
viscosity of the two-phase medium, respectively.
Capillary effects are inserted in the source term ${\bf F}_{st}$ in the form ${\bf F}_{st}=\sigma \kappa {\bf n}_i \delta_i$, where $\sigma$ denotes the surface tension, $\kappa$ is the local mean curvature of the interface, ${\bf n}_i$ is the unit vector normal to the interface and $\delta_i$ is the interface delta function \cite{Brackbill}.\\
The above one-fluid system is the classical model for multiphase incompressible flows with sharp interfaces and effects of uniform surface tension.  Localizing the interface requires solving a transport equation for the characteristic function $C$. This can be performed
 using the volume fraction (also noted $C$ in VOF methods)
 or using a continuous level-set function $\phi$ such that $\phi=0$ at the interface and $\phi>0$ (resp. $<0$) in fluid 2 (resp. 1). In this case, the characteristic function is obtained
 as  $C=H(\phi)$ where $H$ is the Heaviside function\cite{Osher}.\\
 %A specific volume force is also added at the interface to account for surface tensions.
 Note that the above equations are distinct from those involved in various models of
 multiphase turbulence, that may involve Reynolds stresses, terms accounting for unresolved eddies, turbulent viscosities and diffusivities, and non-sharp interfaces. 

\subsection{The Th\'etis code}

Th\'etis is a CFD code developed
in the TREFLE department of the I2M laboratory. It solves the one-fluid Navier-Stokes equations
discretized with implicit finite-volumes on an
irregular staggered Cartesian grid. A second-order centered scheme
is used to approximate the spatial derivatives while a second-order
Euler or Gear scheme is used for time integration \cite{Patankar}. All terms are
written at time $(n+1) \Delta t$, except the inertial term which is
expressed in the following semi-implicit manner:
\begin{equation}
{\bf u}^{n+1} \cdot \nabla {\bf u}^{n+1} \approx \left( 2 {\bf u}^{n} - {\bf u}^{n-1} \right) \cdot
\nabla {\bf u}^{n+1}.
\end{equation}
It has been shown that this approximation allows to reach second-order convergence in time \cite{Poux}.
The coupling between velocity and pressure is ensured by using an
implicit algebraic adaptive augmented Lagrangian method \cite{Vincent6}. The
augmented Lagrangian methods used in this work are independent
of the chosen discretization and could for instance be implemented in
a finite-element framework \cite{Bertrand}. In two dimensions, the
standard augmented Lagrangian approach \cite{Fortin} can be used to
deal with two-phase flows since direct solvers \cite{Pardiso} are efficient in this
case. However, as soon as three-dimensional problems are
considered, the linear system resulting from the discretization
of the augmented Lagrangian terms has to be treated with a BiCG-Stab
II solver, preconditioned by a Modified and
Incomplete LU method \cite{Vorst}. As for the interface tracking and advection of $C$, two different volume of fluid (VOF) methods have been implemented in Th\'etis \cite{Youngs} \cite{VincentVOFSM}. They are evaluated here. The above numerical methods and the one-fluid model have been validated in previous works, \textit{e.g.} \cite{Randrian1} and
\cite{Vincent2}.

\subsection{The Gerris Flow Solver code}

The Gerris Flow Solver (GFS) is an open source code implementing finite volume solvers on an octree adaptive grid together with a piecewise
linear VOF interface-tracking method \cite{popinet03,Popinet}.
{The variables are located on the octree grid in a collocated manner.
Time-advancement is
achieved through a Bell-Collela-Glaz scheme for advection
and a Crank-Nicolson algorithm for viscous stresses. Incompressibility
is satisfied at the end of each time step through a projection method, with
a second MAC projection of the face velocities. The resulting code is second-order accurate in
both time and space for single-phase flows.}
The simulations reported in this paper use a VOF method of the piecewise linear type, in which the interface segments are reconstructed using the mixed-Youngs-centered (MYC) approximation \cite{aulisa}. While not the most accurate for very fine grids, the MYC approximation is easily implemented when using only information from the nearest-neighbour cells, an important advantage when domain-decomposition on octrees is used. Advection of $C$ is performed using the Lagrangian-Explicit or ``CIAM'' method first published by Li \cite{li} and discussed in \cite{scardo,aulisa2,Tryggvason}. While not volume-conserving to machine accuracy, it has very good volume and mass conservation properties.
Surface tension is a vexing question in multiphase flow simulations. As density and viscosity ratios become very large or very
small, the simulations become increasingly difficult with standard methods \cite{Tryggvason}. The problem has been considerably improved in Gerris
as a result of the use of Height-Function methods \cite{Popinet}
and a so-called balanced-force algorithm \cite{francois,renardy:02}. For a review
of surface-tension methods including a discussion of the differences between those used in
Gerris and in other codes, see \cite{popinet2018numerical}. Gerris has been used with success in two-dimensional \cite{fuster13} and three-dimensional \cite{fuster09a,fuster09b} atomization and droplet impact studies. 

\subsection{The Jadim code}\label{jadimsection}

JADIM is a versatile code developed for a number of years at IMFT.
In JADIM, the %time evolution of two-phase or three-phase flows is obtained using the one-fluid formulation of the Navier-Stokes equations. The 
momentum equations are discretized on a staggered
orthogonal grid using a finite-volume approach. Spatial discretization
is performed using second-order centered differences. Time-advancement is
achieved through a third-order Runge-Kutta algorithm for advection/source
terms and a Crank-Nicolson algorithm for viscous stresses. Incompressibility
is satisfied at the end of each time step through a projection method. More
details may be found in \cite{Calmet}. The resulting code is second-order accurate in
both time and space for single-phase flows.

In two- and three-phase configurations, a VOF method with no interface reconstruction is used.
The advection equation for $C$ is solved using a \red{Zalesak} flux-corrected scheme \cite{Zalesak} split into successive one-dimensional steps  \cite{Bonometti} \red{in order to avoid the well-known tendency of the multidimensional version of this scheme to distort iso-$C$ surfaces.}
Owing to the splitting procedure, the overall transport scheme is not rigorously conservative. Therefore a local mass error control which improves upon the global control technique described in  \cite{Bonometti} is employed here.
%
%% The advection equation for $C$ is solved using a flux-corrected scheme split into successive
%% one-dimensional steps \cite{Bonometti}.
%% Owing to the splitting procedure, the overall transport
%% scheme is not rigorously conservative. Therefore a local mass error
%% control which improves upon the global control technique described in [56] is
%% employed.
%% %
The corresponding strategy is based on the detection of drop or
bubbles (through an index function and a topological monitoring of each
bubble or drop) and on an iterative solution which modifies the volume
fraction within the transition regions (\textit{i.e.} those in which $0<C<1$) in order to keep the volume of each
bubble or drop constant.
Since no interface reconstruction step is involved, smearing of interfaces
frequently occurs in high-shear regions.
\red{The original strategy described in   \cite{Bonometti}   is employed to keep this smearing within reasonable bounds.} 
%
%% A specific strategy is employed to keep this smearing
%% within reasonable bounds. For this purpose, the velocity at nodes crossed by
%% the interface is modified to keep the thickness of the interfacial region constant.
%% Details may be found in \cite{Benkenida,Bonometti}.
%% %
In the most difficult cases (\textit{e.g.} break-up
and coalescence), this approach is supplemented by an antidiffusion
technique in which the local volume is redistributed in the direction normal
to the interface so as to eliminate spurious values of the volume fraction outside
interfacial regions.

The capillary force is represented using a modified version of the Continuum
Surface Force (CSF) model \cite{Brackbill}.
This modification consists in defining the interfacial region as that in which the capillary force takes non-zero values rather than that in which the volume fraction takes intermediate values. By doing so, it improves the control of the thickness of the transition region between the two fluids. 
\red{The above strategy for the transport of $C$ makes the approach selected in JADIM intermediate between VOF and Level Set techniques: it is based on the transport
of the local volume fraction of one of the fluids but this transport is not achieved in a strictly conservative manner and no explicit reconstruction of the interface is carried out, the position of the interface being merely identified \textit{a posteriori} with the $C=0.5$ iso-surface.}
%
%% This modification consists in
%% an original manner of evaluating the interfacial area.
%% It allows us to define the interfacial
%% region as that in which the capillary force takes non-zero values (not
%% as that in which the volume fraction takes intermediate values) and improves
%% the control of the thickness of the transition region.\\
%% %

\subsection{The Archer code}\label{archersection}

Archer is a CFD code developed at CORIA laboratory, mainly devoted to multiphase flows. It was previously applied to compute atomization \cite{Menard}, vaporization and mixing \cite{Duret2013130}, and other complex interfacial flows \cite{Aniszewski201452}.
The Level Set (LS) method \cite{Osher} is used for tracking interfaces and shapes. It is based on a continuous distance function $\phi$ defined as the signed distance between any point of the domain and the interface. Similar to the volume fraction $C$, $\phi$ obeys a pure advection equation.

%Motion of the interface in a given velocity field \textbf{u} reads :

%\begin{equation}
%\label{levelset}
  %\frac{\partial \phi}{\partial t}+ {\bf u} \cdot \nabla \phi = 0
%\end{equation}

To avoid singularities in the $\phi$ field, the fifth-order conservative WENO \cite{shu} scheme is applied to discretize convective terms. When the LS advection is carried out, high velocity gradients can cause wide spreading or stretching of $\phi$ which then no longer remains a distance function. A redistancing algorithm \cite{sussman98} is thus applied at every time step to restore the distance property of $\phi$, \textit{i.e.} $|\nabla\phi|=1$. Advancement of the $\phi$-equation and the redistancing algorithm can induce mass loss in under$-$resolved regions  \cite{Aniszewski201452}. This is the main drawback of LS methods. To improve mass conservation, the Coupled Level-Set Volume of Fluid (CLSVOF) method \cite{sussman20} is used. The main idea of this approach is to benefit from the advantages of each tracking strategy: minimize the mass loss using VOF and keep a fine description of interface properties with the smooth LS function. (A study of the performance of various VOF variants was recently published \cite{Aniszewski201452} using Archer in the CLSVOF context; the phase inversion problem is also discussed therein.) The coupling between LS and VOF is maintained using a correction scheme based on a geometric reconstruction of both VOF and LS interfaces \cite{Wojtek}. The LS method itself is coupled with a projection method for the incompressible Navier-Stokes equations, where the density and the viscosity depend on the sign of the LS function (with appropriate interpolations used in interfacial cells). To finalize the description of the two-phase flow, jump conditions across the interface are taken into account using the Ghost Fluid (GF) approach. In this approach, ghost cells are defined on each side of the interface \cite{kang} \cite{liu} and appropriate discretization schemes are applied to the jump of each quantity. As defined above, the interface is characterized through the distance function, and jump conditions are extrapolated on some nodes on each side of the interface. Following the jump conditions, the discontinued functions are extended continuously and then derivatives are estimated. Heaviside functions are designed according to the distance function in order to provide a characteristic function for the two-phase medium. They follow the functions proposed in formula (6) of \cite{Towers} . Discrete delta functions are built in a similar way and are regularized using a properly set cosine function. These techniques have been presented and validated in previous works, \textit{e.g.} \cite{Menard}.

\subsection{The DyJeAT code}

DyJeAT( Dynamics of Jet ATomisation) is an in-house computational fluid dynamics library developped at ONERA to deal with atomization processes in aircraft or launcher engines. It is based on the one-fluid formulation for incompressible two-phase flows with specific interface capturing features.
Classical projection methods are used to ensure the incompressibility constraint \cite{chorin,temam}. The spatial discretization is based on staggered uniform Cartesian grids for the velocity components, all others quantities such as density, pressure and level-set function $\phi$ being cell-centered. Advection terms in the momentum equations  are approximated in a conservative way with $5^{th}$-order accurate WENO schemes \cite{shu}. This particular choice was motivated by the robustness and low numerical dissipation of these schemes observed in DNS studies \cite{Trontin}. Viscous terms are discretized with second-order centered scheme. Time integration is performed with a second-order accurate Adams-Bahsforth scheme. The Poisson equation for the pressure is solved by a fast multigrid preconditioned conjugate gradient method \cite{tatebe}.
A level-set method \cite{Osher} is used to capture the interface, implicitly defined as the zero level of the smooth function $\phi$. %Moreover, $\displaystyle{\phi}$ is imposed to be the signed distance to the interface. %This property ensures that the level-set function is well behaved at the interface between the two fluids. 
Similar to the momentum equation, the level-set equation is solved with a $5^{th}$-order  WENO scheme for the spatial discretization and a $3^{rd}$-order TVD Runge-Kutta scheme for time advancement.
%
%%During its transport by the fluid velocity, the level-set function  no longer remains a distance function ($\displaystyle{ |\nabla \phi| \neq 1}$), which  can lead to large mass losses or gains. Consequently, the level-set must be repeatedly reinitialized. This is achieved by solving to the steady state the Hamilton-Jacobi equation for $\displaystyle{\phi}$ \cite{sussman}. We use a  high-order WENO5-Godunov scheme \cite{jiang} for the discretisation of the Hamilton-Jacobi equation.
%
\red{During its transport by the fluid velocity, the level-set function no longer remains a distance function.  Since redistancing is mandatory for curvature computation, and as is standard practice, the level-set implicit function is repeatedly reinitialized to the signed distance. Redistancing in the level-set scheme, including the associated resolution issues and mass conservation properties, is discussed in detail in \cite{couderc:tel-00143709} (see in particular figures 3.8 and 3.9 therein).}
Jump conditions at the interface for both pressure and viscous terms are handled with a Ghost Fluid approach \cite{kang,liu}.

\subsection{\textcolor{black}{Pros and cons of the different codes}}

\begin{table}[ht]
\begin{center}
\begin{tabular}{cccccc}
\hline \hline
 Code & Interface & Navier-Stokes  & Capillary & Grids & Solvers\\
  & tracking & velocity-pressure  & forces & &  \\
  &  & coupling  &  & &  \\
\hline\addlinespace[.15cm]
Archer & CLSVOF & Projection & Ghost & Cartesian & BiCGStab \\
 &  & method & fluid & staggered & Multigrid \\ \addlinespace[.15cm]
DyJeAT & Level set & Projection & Ghost & Cartesian & BiCGStab \\
 & & method & fluid & staggered & Multigrid \\\addlinespace[.15cm]
Gerris & Geometrical & Projection & CSF with & Collocated & BiCGStab \\
 & VOF & method & height functions & and AMR & Multigrid \\\addlinespace[.15cm]
Jadim & Implicit & Projection & CSF & Cartesian & BiCGStab \\
 & VOF & method &  & staggered & Multigrid \\\addlinespace[.15cm]
Thetis & Geometrical & Augmented & CSF with & Cartesian & BiCGStab \\
 & VOF & Lagrangian & smooth VOF & Staggered & ILU \\\addlinespace[.1cm]
\hline\hline
\end{tabular}
\end{center}
\caption{\textcolor{black}{Summary of the numerical methods used in the various codes.}}\label{tabCode}
\end{table}

  {All the codes used in the present work are based on
  finite volumes but with different variants of the methods.
  Obviously, each code can lead to a different
  result on a given grid, due to variations in the methods used.
  Table \ref{tabCode} is given to clarify the comparison of
  the model and numerical methods used. Schematically, the codes cover
  the broad range of numerical methods that can be found in the
  literature for simulating multiphase flows using the sharp interface
  approximation described above. This approach is sometimes called
  direct or detailed, to contrast it with other approaches that
  depart from the sharp interface approximation.
  Four of the codes use staggered grids while one (Gerris) is based on
  collocated unknowns. Staggered grids have several advantages: a more accurate
  pressure solution, and the avoidance of spurious ``red-black'' velocity oscillations.
  However, Gerris has demonstrated high accuracy on test cases \cite{Popinet,fuster09b}
  and ten years of experience with Gerris show almost no occurrence of
  spurious oscillations. 
  Four codes are using a projection technique achieving a velocity-pressure coupling that contains a time
  splitting error, while one code (Thetis) uses an exact augmented Lagrangian
  approach. This technique is exact if the residual of the iterative
  solver is zero at machine accuracy. In terms of interface
  tracking, Gerris and Thetis are using a geometrical VOF-PLIC
  technique while Jadim employs a FCT scheme to directly approximate
  the advection of the volume fraction without reconstructing explicitly the interface geometry (this is why the corresponding approach may be thought of as Implicit VOF).
  % The FCT-VOF does not preserve mass while VOF-PLIC does.
  DyJeAT uses the Level Set method
  whereas Archer couples this technique with VOF-PLIC in order to improve
  mass conservation and also to better evaluate advective effects in
  the momentum transport. Concerning capillary effects, the Ghost Fluid approach based on jump
  relations at the interface and the CSF formulation with height functions \cite{Popinet} are the
  most accurate. Thetis
  and Jadim use less sophisticated capillary force approximations
  that are known to generate larger spurious currents than the other
  methods. To finish with numerical methods, all projection steps
  {except for Gerris} are achieved with an iterative BiCGStab II solver preconditioned with a
  multigrid algorithm, while Thetis and the coupled augmented
  Lagrangian method use BiCGStab II and an incomplete LU
  preconditioner because the linear system is not symmetric. The major drawback of
  the Thetis solver is that its parallelization has a good speed-up
  until $1000$ processors but can hardly handle in its present form
  a larger number of processors due to the ILU preconditioner. In contrast, the
  multigrid preconditioner used in the projection approach is nicely extendable in
  parallel computations until $100,000$ processors. {In Gerris, the projection step is achieved thanks to an in-code multigrid Poisson solver.}} \\

%\textcolor{black}{On a general point of view, its difficult to give detailed reasons or scientific explanations on why similar models and numerical methods can provide different results on the same problem, as observed for high order moments of velocity or interface in the next sections of the paper. In fact, a first reason is linked to the unsteady and turbulent character of the considered flows that amplify numerical errors differently with different numerical techniques and solvers. At the end, they lead do different instability growth and turbulent character of the flow, and so for associated interfacial structures. A second reason is that the field of multiscale multiphase flow is still not well understood, in particular with respect to unsteadiness and turbulence, enstrophy statistics or fragmentation events. This is the reason why the phase inversion benchmarks have been designed and chosen,{\it i.e.} to try to determine the limits of existing multiphase flow codes.}

%%================================================================================

% Part 1: Results using DYJEAT 

%%================================================================================

\section{Detailed numerical simulations using DyJeAT}

{We used the DyJeAT code to obtain results at very high resolution. The grids we used are still too coarse to reach the Kolmogorov scale but can be considered suitable to provide ``detailed numerical simulations'' offering a view of the large-scale vortical structures. The meaning of such detailed simulations for interfacial structures may be understood from a study of the convergence of the flow properties. As we shall see, some of the characteristics of the flow linked to its multiphase character, such as the PDF of droplet sizes, also seem to converge. We can thus consider that the detailed numerical simulations also offer a view of the large-scale properties of the interfacial structure. As a preliminary result relevant to both cases, we show in Figure
\ref{en_both} the evolution of the total mechanical energy
  $$
  E_{m} = E_{p,1} + E_{k,1} + E_{p,2} + E_{k,2} \,,
  $$
  \begin{figure}[ht!]
\begin{center}
  \includegraphics[width=6cm]{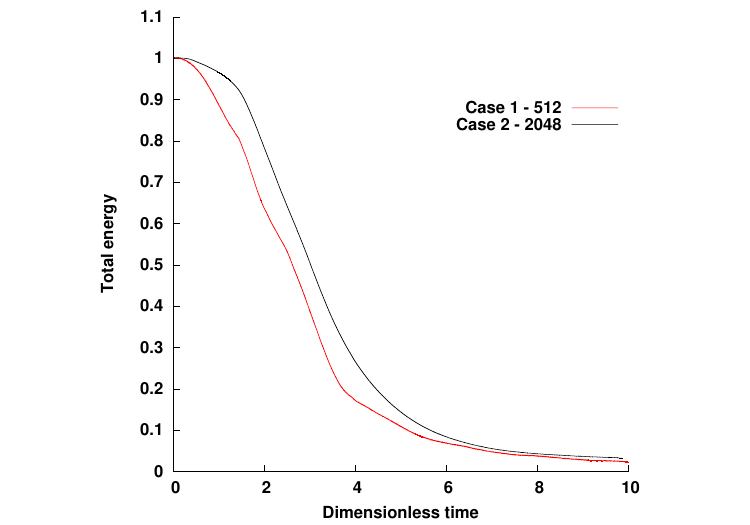}
  \end{center}
\caption {Evolution of the rescaled  (see text) total mechanical energy for both cases using DyJeAT. }
\label{en_both}
\end{figure}
  obtained on a $512^3$ grid in both cases which, as we shall see, are well converged.
  The initial energy is $E_m(0) = E_p^i$. As $t\rightarrow \infty$ one expects the system to converge to a state
  in which the light fluid occupies exactly the cuboid $\Omega_{up}$ and the kinetic energy vanishes, so
  we expect $E_{m} \rightarrow E_m^f = E_p^f$. We rescale $E_m(t)$ for both cases as
  $$
  E_m^{'}(t) = \frac{E_m(t) - E_p^i}{E_p^f - E_p^i} \,.
  $$
  The plot shows a remarkable similarity in both cases, especially at the end of the process where the two curves nearly superpose. Energy in case 1 is somewhat lower, presumably because a fraction of it was transferred to surface energy. For the same reason, one observes more fluctuations in the curve corresponding to case 1. The similarity between the two cases is an indication of the possibility of the \Re- \mbox{} and \We-independent fully developed turbulent regime discussed in the introduction. }

\subsection{Case 1: a phase inversion problem with few fragmentation events}

In this section we present the numerical results pertaining to case
1 obtained on
$128^3$, $256^3$ and $512^3$ grids with DyJeAT. The characteristics of this case are summarized in Tables
\ref{tab:dim}-\ref{tab:nondim}.  Dimensionless data are considered
according to the definitions indicated in Table \ref{tab4.1}. 

\begin{figure}[ht!]
\begin{center}
  \includegraphics[width=4cm]{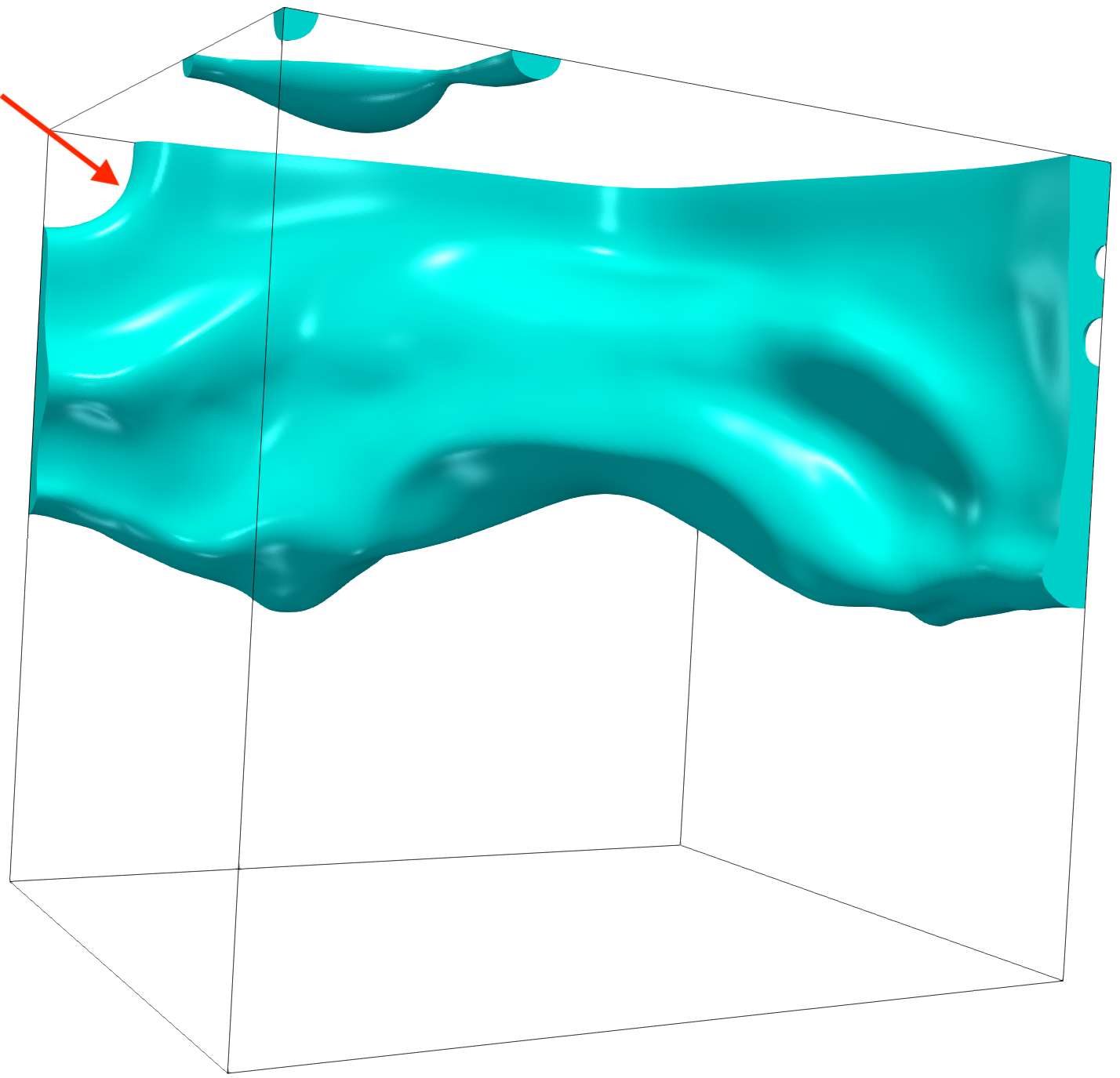}\hskip 10pt \includegraphics[width=4cm]{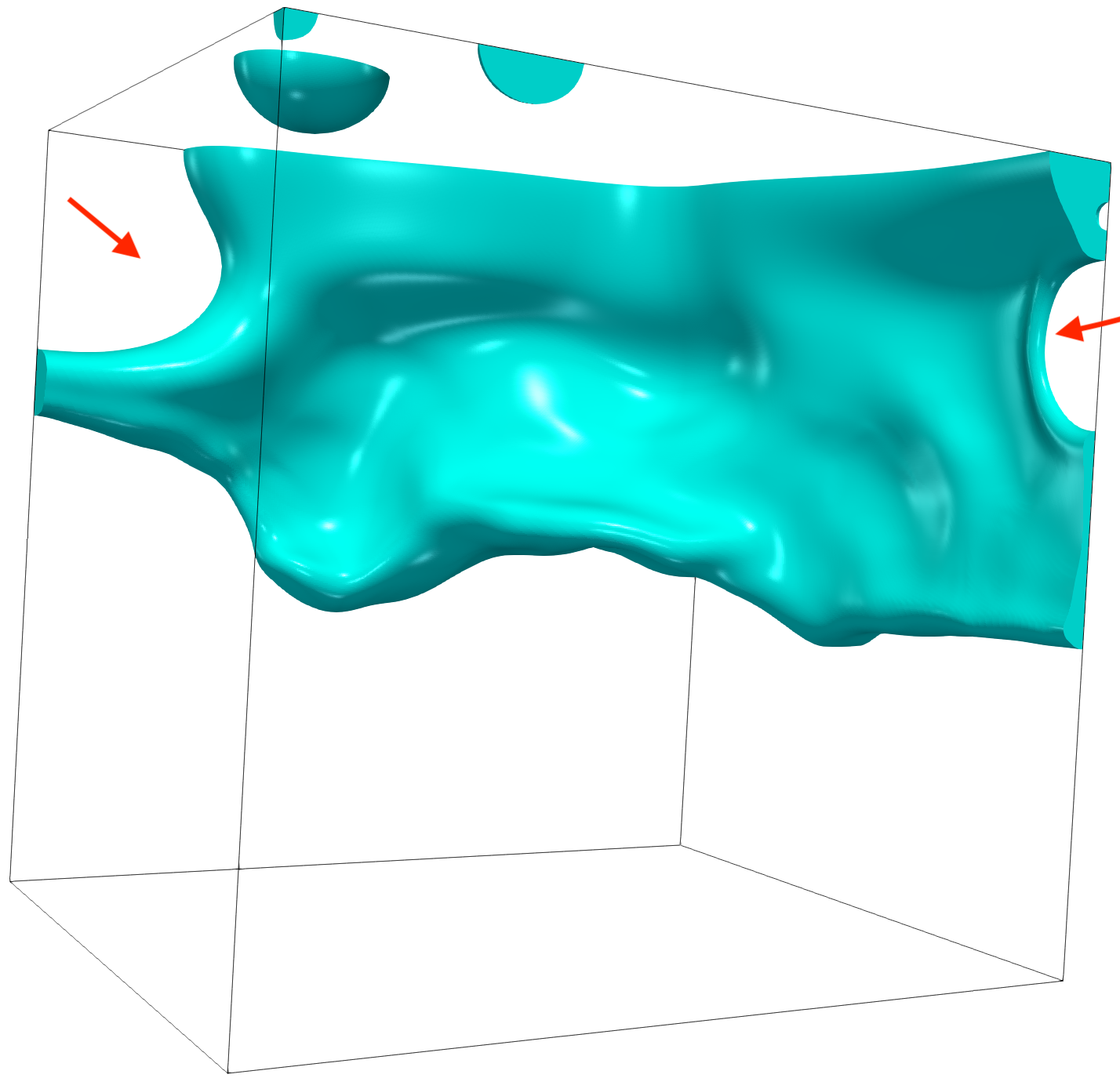}\hskip 10pt   \includegraphics[width=4cm]{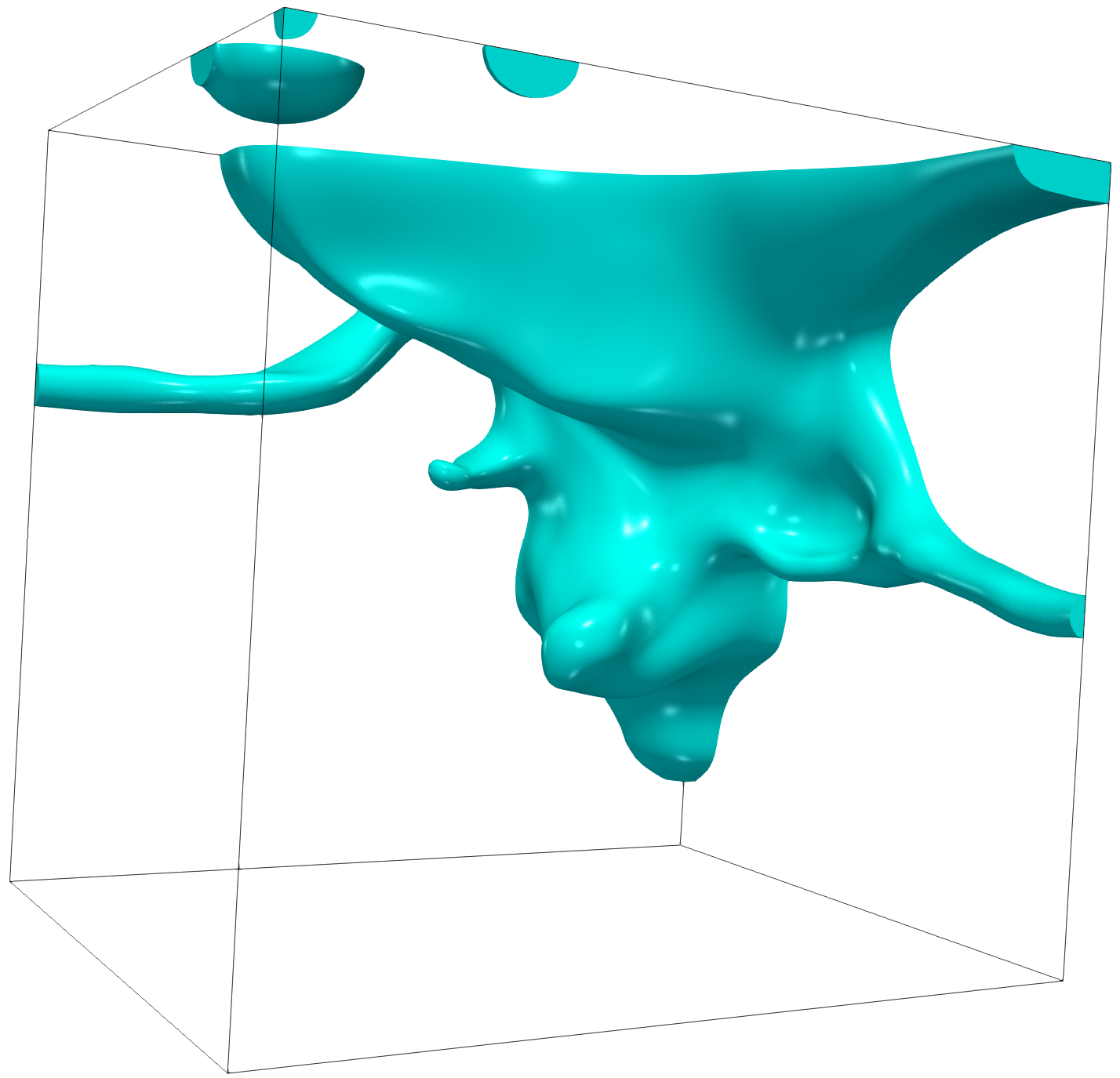}
  \end{center}
\caption {A sequence of interface shapes for case 1, using DyJeAT at $512^3$ resolution. The dimensionless times are (from left to right)
  $t^*=2$, $t^*=2.25$ and $t^*=2.5$. One can see a large sheet pierced by
  two expanding holes (arrows).}
\label{int_c1}
\end{figure}

{A representation of a sequence of interfacial shapes is shown in Figure \ref{int_c1}.
  It is seen that the interface forms a thin sheet with expanding holes that lead to the formation of
  ligaments and then droplets. The expanding holes are surrounded by a characteristic Taylor-Culick rim \cite{taylor59c,Culick60}.}

\begin{figure}[ht!]
\begin{center}
  \includegraphics[width=8cm]{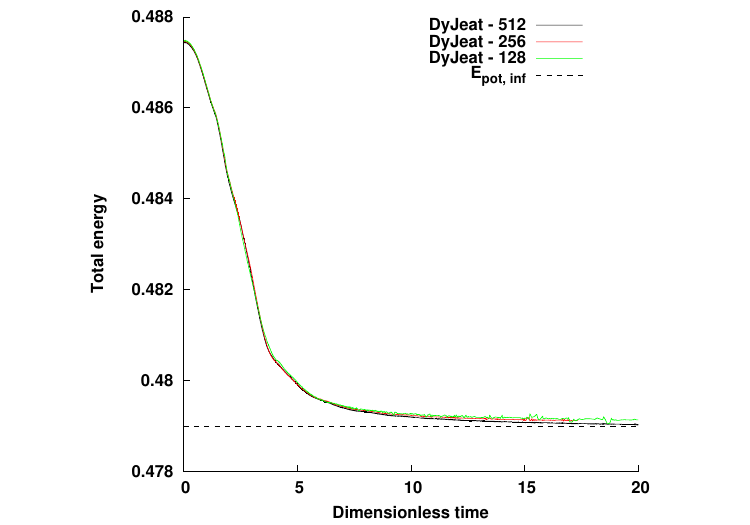}
  \end{center}
\caption {Evolution of the dimensional total mechanical energy (without the surface energy), $E_m$, as a function of dimensionless time for case 1.
The horizontal line is the expected value $E^f_{p}$. }
\label{em}
\end{figure}

{The potential and kinetic energies are mostly dependent on the large scales of the flow and are thus
  ``macroscopic'' quantities which, by definition, are related to the large scales and should for that reason
  be well resolved on moderately fine grids. 
 \red{ A simple verification that this is indeed the case is achieved by
  plotting the total mechanical energy
  as done in Figure \ref{em} for case 1. It is seen that the energy evolves in a nearly monotonic manner, and that 
  it asymptotes to the expected value $E^f_{p}$ at long times.}}

\clearpage

\begin{figure}[ht!]
\begin{center}
  \includegraphics[width=6cm]{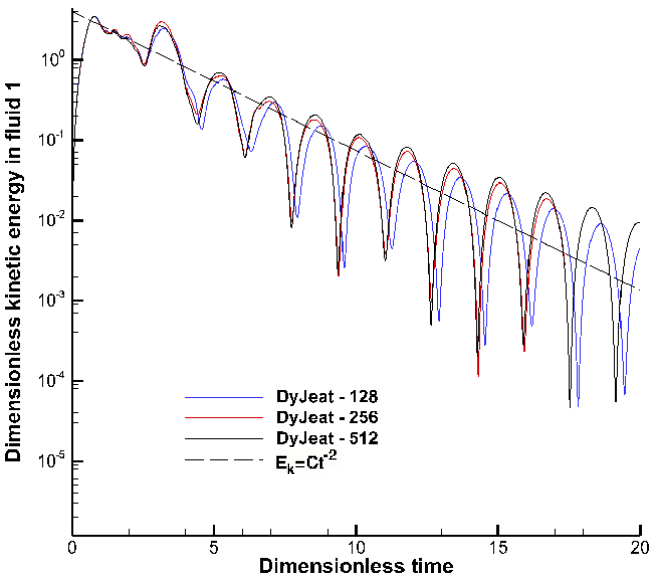}
  \includegraphics[width=6cm]{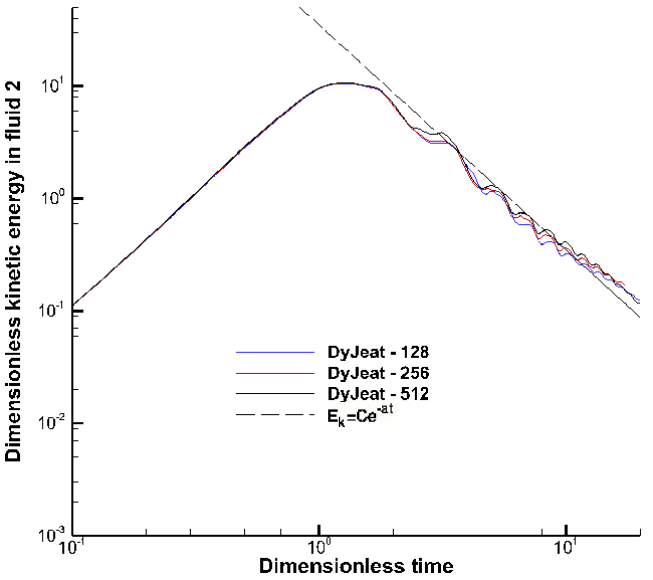}
  \end{center}
\caption {Grid convergence of kinetic energies for case 1.}
\label{fig4.3}
\end{figure}

\begin{figure}[ht!]
\begin{center}
  \includegraphics[width=6cm]{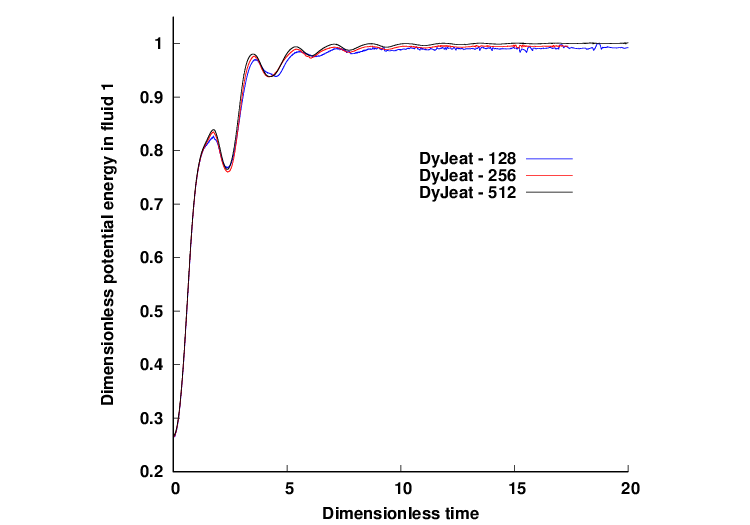}\hskip 7pt
  \includegraphics[width=6cm]{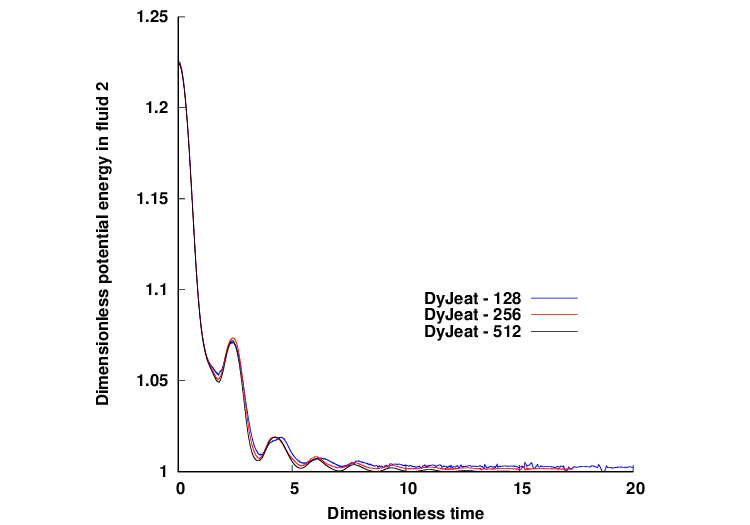} \hskip 7pt
  \end{center}
\caption {Grid convergence of potential energies for case 1.}
\label{fig-Epot-case1}
\end{figure}

  %A convergence study for the potential and kinetic energies was carried out on $128^3$, $256^3$ and $512^3$ grids with the DyJeAT code. 
  The corresponding results for the kinetic and potential energies are reported in
Figures \ref{fig4.3} and \ref{fig-Epot-case1}, respectively.
%There is still some difference between the 256 grid and the 512 grid
There is still some difference between the $256^3$ grid and the $512^3$ grid
for fluid 2 around time $t^* ={2.5}$, although not in fluid 1.
%the fluid 2 case around time 25, although not in the fluid 1 case.
{We note that, since the kinetic energy is scaled by
${(1/16) \rho_n U_g^2 H^3}$ for fluid $n$ (see Table \ref{tab4.1}), the maximum kinetic energy of fluid 2
in the scaled variables is $8 (U_K/U_g)^2 (\rho_1/\rho_2) = 11.1375$. This is close to the maximum
of the curve for fluid 2 in Figure \ref{fig4.3}. The fact that the kinetic energy is close to its upper bound also
means that $U_K$ is close to the $L_2$-norm velocity, and that our Reynolds and Weber number estimates are
consistent with the observed $L_2$-norm velocity.}

%% To characterize the phase separation from a macroscopic point of view, the volume ratio of fluid 1 occupying the top $H/8$ part of the box is plotted in Figure \ref{fig4.4dyjeat}.  A convergence study was carried out on
%% $128^3$, $256^3$ and $512^3$ grids. This time the difference between the $256^3$ grid and the $512^3$ grid
%% is much smaller than the difference between the $256^3$ grid and the $128^3$ grid, indicating convergence.
%% We note that the effective initial volume of light fluid is not strictly equal to the theoretical volume of $H^3/8$,
%% and that in some cases one observes liquid drops trapped on the domain walls. 
%% Both of these effects explain the fact that numerically obtained volume ratios tend to a value slightly less than 1.
%% This plot also suggests that phase separation is almost completed approximately $20$ time units, which is approximately twice longer than the ten time units it takes for the potential energy to equilibrates (Figure \ref{fig-Epot-case1}). This may be explained as follows. After $t^*=10$
%% most of the fluid is near the top of the box. The oil-water interface is {\em on average} at its equilibrium position, but is affected by gravity waves, and thus keeps oscilating as seen on Figure \ref{fig4.4dyjeat}.

\begin{table}[ht]
\begin{center}
\begin{tabular}{ccc}
\hline
 Parameter & Value & Units \\
\hline
 $t^*={t}/{t_c}$ & $\displaystyle{\frac{t}{0.643}}$ & - \\
$E_{p,1}^f$ the potential energy in fluid 1 for $t \rightarrow  \infty$ & $0.1035$ & J \\
$E_{p,2}^f$ the potential energy in fluid 2 for $t \rightarrow  \infty$ & $0.3755$ & J \\
$\displaystyle{E_{k,1}^*=\frac{E_{k,1}}{(1/16) \rho_1 U_g^2 H^3}}$ & $\displaystyle{\frac{E_{k,1}}{0.000341}}$ & - \\
$\displaystyle{E_{k,2}^*=\frac{E_{k,2}}{(1/16) \rho_2 U_g^2 H^3}}$ & $\displaystyle{\frac{E_{k,2}}{0.000378}}$ & - \\
${{H^3}/{8}}$ the final volume of fluid 1 in the top part of the box & $0.000125$ & $m^3$ \\
Maximum of enstrophy in fluid 1 (DyJeAT on a $512^3$ grid) & $0.0733$ & $m^{3}.s^{-2}$ \\
Maximum of enstrophy in fluid 2 (DyJeAT on a $512^3$ grid) & $1.3759$ & $m^{3}.s^{-2}$ \\
\hline
\end{tabular}
\end{center}
\caption{Parameters used to define the dimensionless variables in case 1.}\label{tab4.1}
\end{table}

%% \begin{figure}[ht!]
%% \begin{center}
%%   \includegraphics[width=6cm]{Cas3_EVolFluid1b.png}
%%   \end{center}
%% \caption {Evolution of the volume ratio of fluid 1 in the top part of the box in case 1 reference case).}
%% \label{fig4.4dyjeat}
%% \end{figure}

\begin{figure}[ht!]
\begin{center}
  \includegraphics[width=6cm]{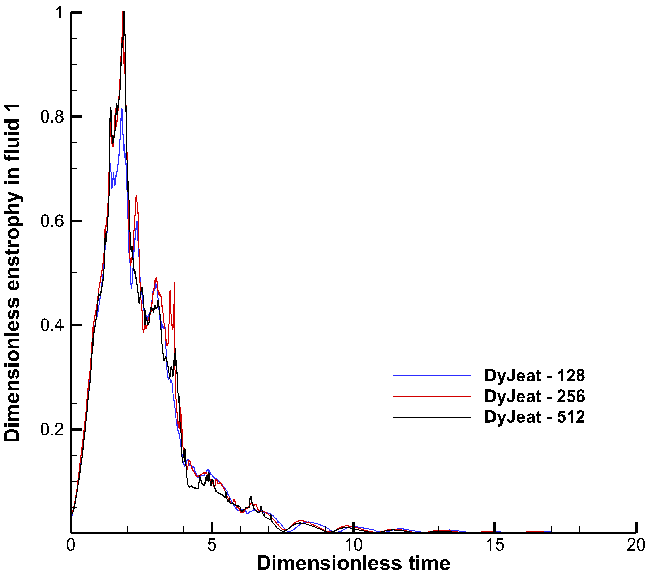}
  \includegraphics[width=6cm]{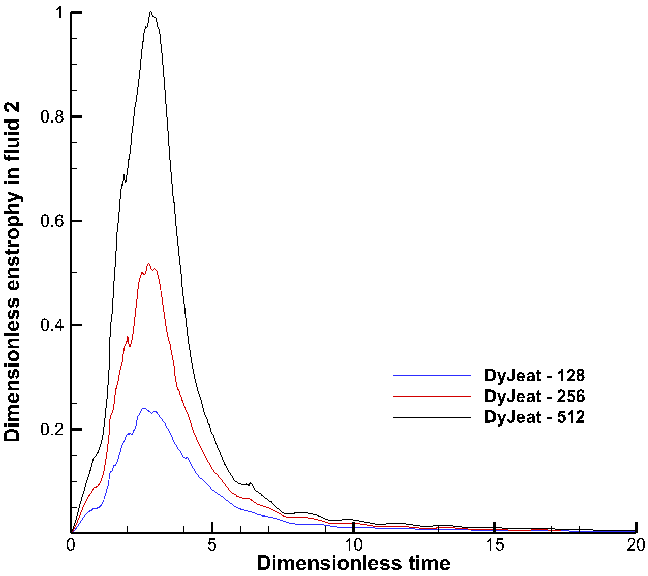}
  \end{center}
\caption {Grid convergence of enstrophy for case 1.}
\label{fig4.6}
\end{figure}

\begin{figure}[ht!]
\begin{center}
  \includegraphics[width=6cm]{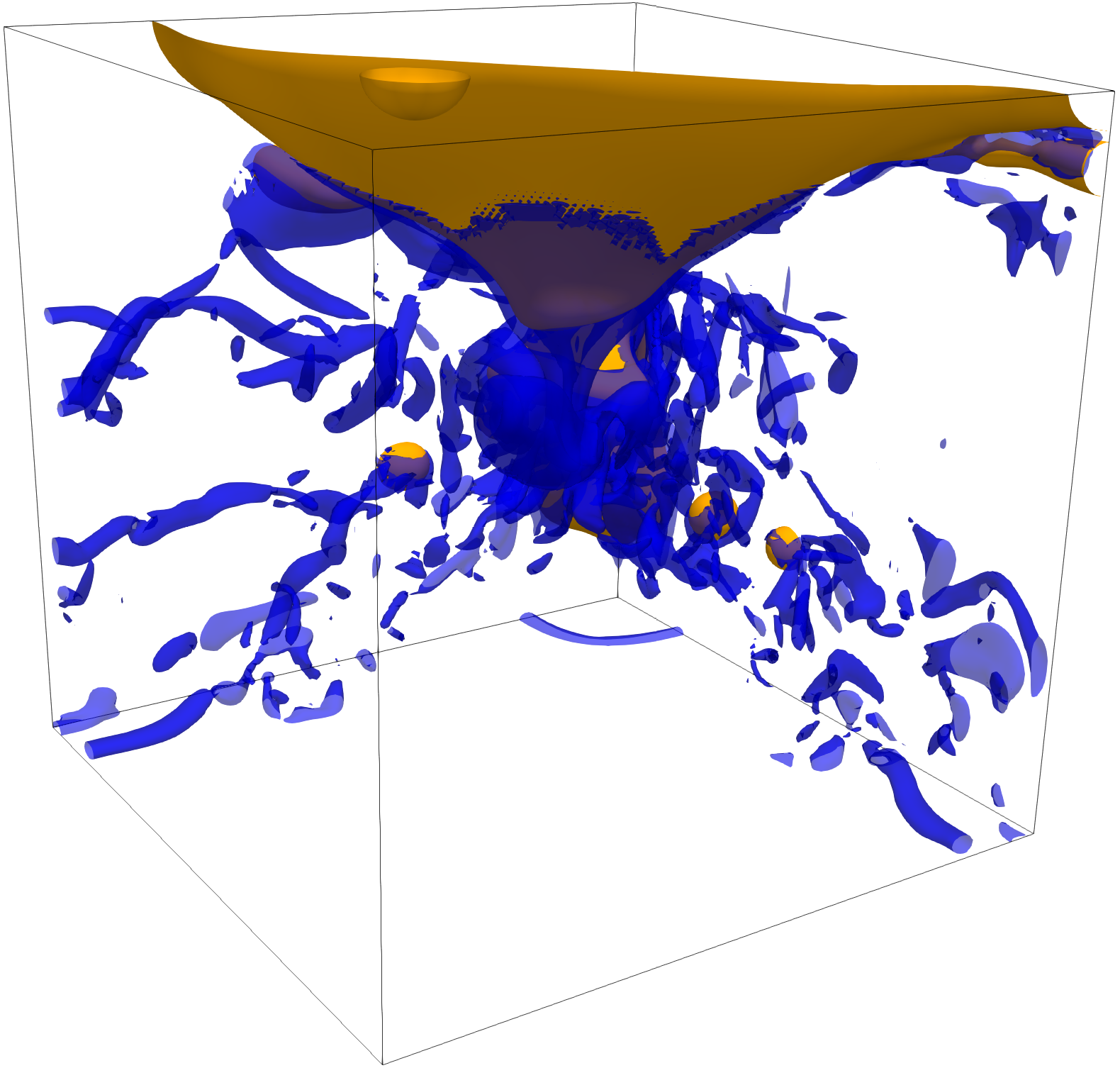}\hskip 7pt
  \includegraphics[width=6cm]{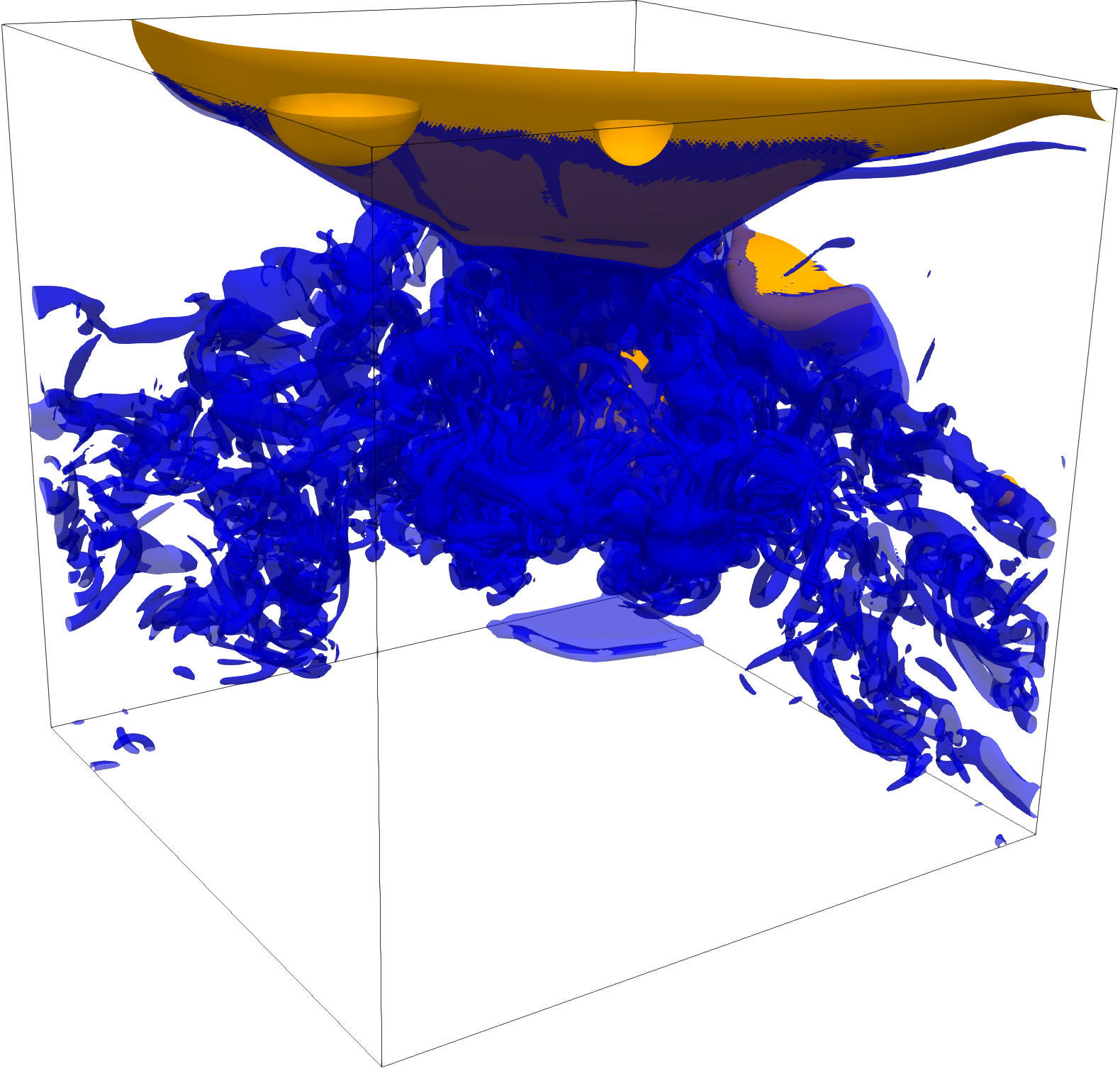} \\
  \end{center}
\caption {Vorticity structures and interface shape for case 1: iso-60 vorticity magnitude (blue) and iso $C=0.5$ interface (orange surface) obtained with DyJeAT on the 128$^3$ grid (left) and 256$^3$ grid (right).}
\label{fig4.6dyjeat}
\end{figure}

The grid convergence results for the enstrophy are illustrated in Figure \ref{fig4.6}. Although the enstrophy seems to converge in fluid 1, its peak increases with the grid resolution in fluid 2 and no convergence is reached even on the $512^3$ grid. Therefore, although primary moments such as potential and kinetic energies or volume ratio of the light fluid suggest that a real DNS was achieved, the analysis of higher-order moments such as enstrophy invalidates this hope.

The interface shape and  isovalues of vorticity magnitude at $t^{*}=3$ (which corresponds to the peak within fluid 2 in Figure \ref{fig4.6}) % \ref{fig4.5} and \ref{fig4.5b}) 
are reported in Figure \ref{fig4.6dyjeat}. %The maximum magnitude of the vorticity is found to be concentrated in the near-wall regions and in regions where the large blob of light fluid that has risen during the phase separation process undergoes high deformations.
Finer vortical structures are clearly captured by the finer grid inside fluid 2 (near the bottom corners for instance). %When the large blob of light fluid moves from one side of the box to the other, heavy fluid is trapped between some of them and the wall, generating high shear (hence high vorticity) regions. Such boundary layers are clearly not fully resolved with the grids used in the present work, which explains the lack of convergence observed for the volume-averaged enstrophy.\\
 %<-It seems that only one code confirms this [WA]
%Contrary to what is observed in bubbly flows, the maximum of vorticity is not restricted to the wake of the droplets, indeed it is present everywhere in fluid 2.
As pointed out in Table \ref{tab:nondim}, the $512^3$ grid does not reach the Kolmogorov scale $\eta_K$ and thus does not resolve the finest vortical structures.
In addition, as we are considering two-phase flows with jumps in the physical properties, especially viscosity, properly capturing the interfacial vortical layers requires that a sufficient number of grid points be located around the interfaces. %The 2D slice shown in Figure \ref{fig4.6b} clearly indicates that the vorticity magnitude reaches its maximum near the interfaces, a region where high shear rates take place, owing to the large viscosity contrast.
Therefore, we may suspect that the way the local viscosity is estimated as a function of the local volume fraction plays a role in the local vorticity magnitude. Indeed, an arithmetic average between $\mu_1$ and $\mu_2$ is generally used but there are good reasons to rather favor an harmonic average \cite{Ritz, PianetMOY, VincentLAG}. To avoid this possible influence, we re-computed case 1\textcolor{black}{, still with DyJeAT,} while considering the same viscosity in both fluids. The corresponding results are discussed in Appendix \ref{AppendixB}. %that's a very nice idea (WA)
{They show that grid convergence of enstrophy is still not achieved when the viscosity is set to $0.1 Pa.s$ (the corresponding Reynolds number is 137) although the enstrophy excursions are less violent.
 A similar study was performed
by some of us \cite{sayadiconvergence} with the conclusion that sheet breakup may be
responsible for spurious large enstrophy even at reduced Re and La.
The obvious conclusion of these
additional computations is that the differences found among the
various enstrophy evolutions are not due to the effects of an inaccurate numerical evaluation of the viscosity jump but to other effects requiring further study. We also note that in
the atomization simulations performed by some of us \cite{ling2019two}, enstrophy as measured statistically is not
diverging upon grid refinement. 
%  whereas it is clearly not achieved with viscosities 10 and 100 times smaller.
%  The obvious conclusion of this specific study is that the Reynolds number selected in the case 1 documented above is far too large for the thin vortical layers to be fully resolved, even on a $512^3$ grid.
Hence the results do not correspond to a true Direct Numerical Simulation: convergence is achieved on quantities dominated by the large-scale motions, such as the kinetic and potential energies, but it is not on enstrophy for which the main contribution is from the small scales.}\\

\subsection{Droplet sizes}

\begin{figure}[ht!]
\begin{center}
  \includegraphics[width=12cm]{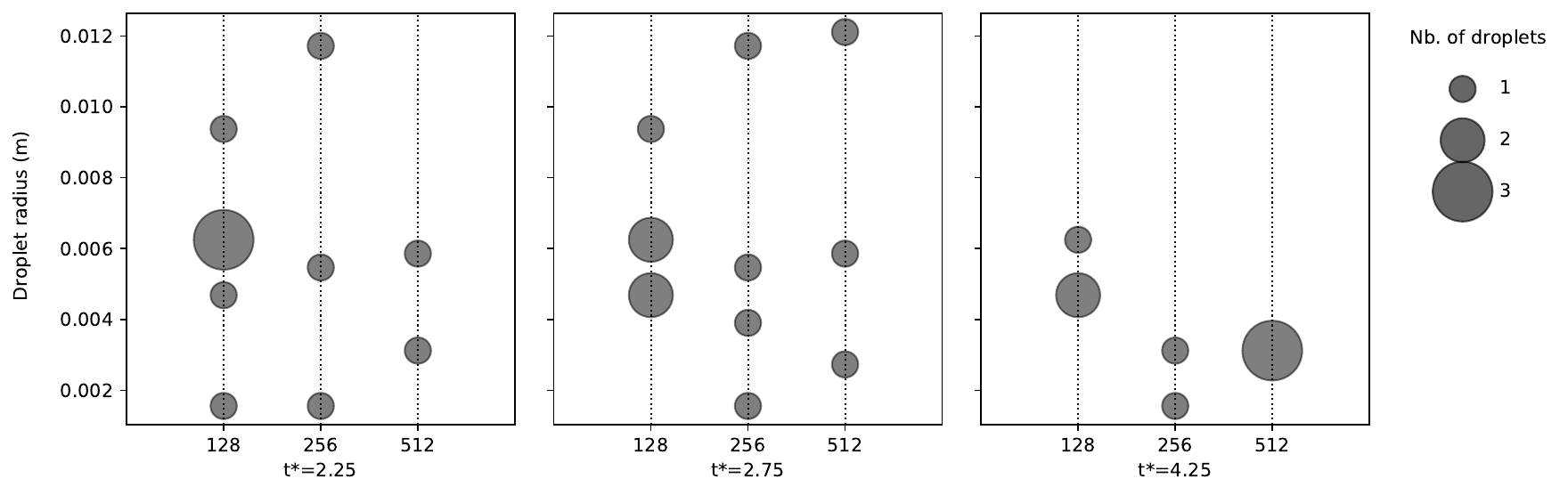}
  \end{center}
\caption {A representation of the sizes of the largest droplets (assuming $H=0.1$\,m) found in case 1, as a function of time and grid resolution. %At later times, no droplets are found. The number of droplets decreases in time as droplets reach the top of the box and merge with the bulk of fluid 1. 
\label{pdf_drops_c1}}
\end{figure}

We now briefly analyze the evolution of the droplets size observed in results for case 1 obtained with DyJeAT. We saw in Figure \ref{int_c1} that the interface forms a thin sheet with expanding holes that lead to the formation of
ligaments and then droplets. This hole formation mechanism was already observed in simulations of atomizing jets \cite{shinjo2010simulation} or mixing layers \cite{ling17} and is critical in the atomization or fragmentation
process \cite{lohse2020double,villermaux2020fragmentation}. It is likely that the control of the hole formation
process is essential for obtaining a convergent droplet size distribution upon grid refinement \cite{leonardo}. \\
\color{red}{Figure \ref{pdf_drops_c1} shows how the distribution of the largest drops observed with DyJeAT varies with the grid resolution and dimensionless time.  At later times, no droplets are found. The number of droplets decreases in time as droplets reach the top of the box and merge with the bulk of fluid 1. Overall, at a given $t^*$, the number of droplets also decreases as the resolution increases. Similarly, for a given resolution, the average size of the droplets decreases as time proceeds.}\color{black}

%----------------

% CASE 2

%----------------

\subsection{Case 2: a phase inversion problem involving many fragmentation events}

In this section we present the numerical results pertaining to case
2. The characteristics of this case are also provided in Tables
\ref{tab:dim}-\ref{tab:nondim}.  Dimensionless data are considered
according to the definitions indicated in Table \ref{tab4.2}. In order to examine grid convergence, the numerical simulations were run on five grids {\it i.e.} $128^3$, $256^3$, $512^3$, $1024^3$ and $2048^3$.
(It is worth noting that over 30 million hours of CPU time were necessary to perform the simulation on the finest $2048^3$ grid.)

\begin{figure}[ht!]
\begin{center}
  \includegraphics[width=8cm]{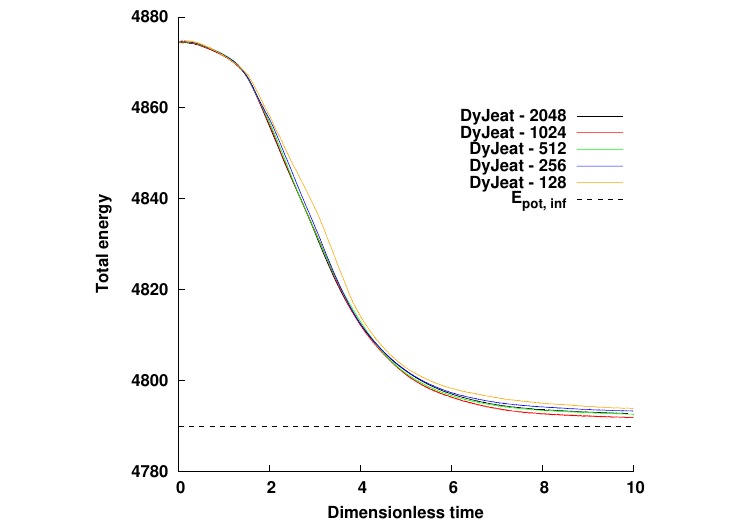}
  \end{center}
\caption {Evolution of the dimensional total mechanical energy (without the surface energy), $E_m$, as a function of dimensionless time.
The horizontal line is the expected value $E^f_{p}$.}
\label{em2}
\end{figure}

{
  \red{A 
  plot of the total mechanical energy is shown in Figure \ref{em2}. As in case 1, the energy decreases monotonically. It is seen that the convergence is irregular, as the $1024^3$ simulation is further away from the reference $2048^3$ simulation than the $512^3$ one. This effect will be seen in more detail in the separate plots for the kinetic and potential energies shown below.  At long times, the computed energy is clearly above  the expected asymptotic value $E^f_{p}$.}
  %% However, we note that
  %% the energy decay is universal as shown in Figure \ref{en_both}. Thus, one can expect
  %% the total energy to continue decaying to $E_p^f$ as in case 1, which was continued for double the dimensionless time
  %% than the present case.
}

\begin{figure}[ht!]
\begin{center}
  \includegraphics[width=6cm]{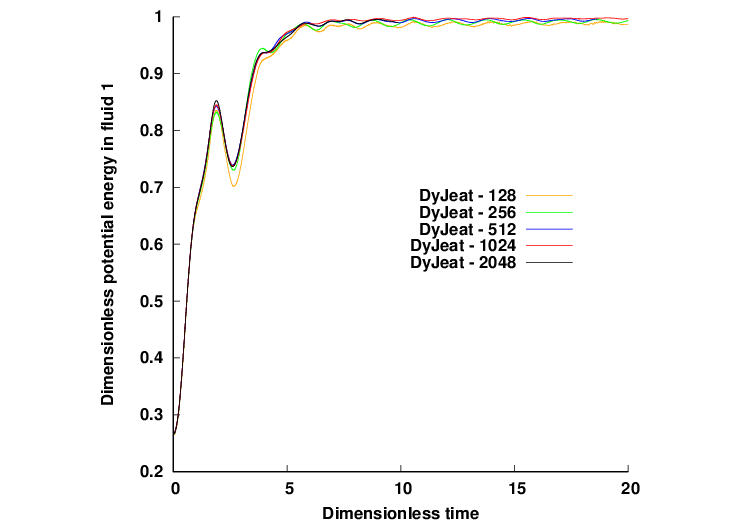}\hskip 7pt
  \includegraphics[width=6cm]{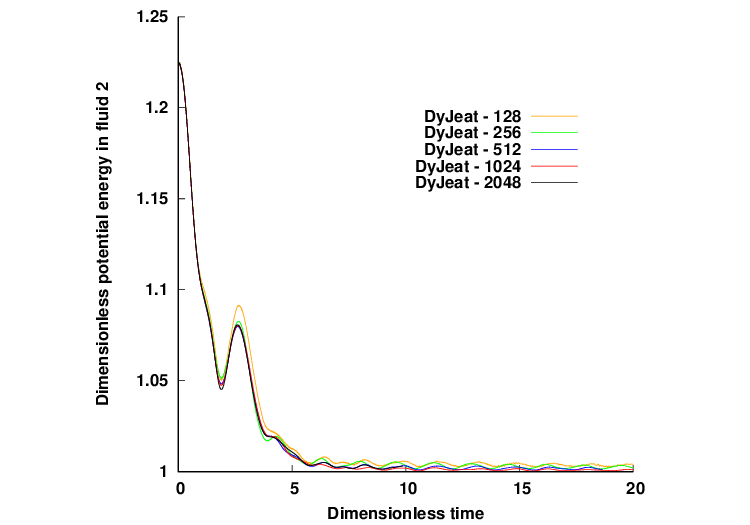}
  \end{center}
\caption {Grid convergence of potential energies for case 2.}
\label{fig-Epot-case2}
\end{figure}

The results for the potential and kinetic energies  are displayed in Figures \ref{fig-Epot-case2} and \ref{fig4.10}, respectively.
The data obtained on the three finest grids superpose until $t^* = 4.5$ approximately.
We characterized the rate of convergence by comparing the potential energies at $t^* = 2.63$. The corresponding conclusions are reported in
Table \ref{tab.conv}. The order of convergence is near $p=2$ for the low-resolution grids and near $p=1$ for the high-resolution ones. It is likely that at low resolution, the error on the flow characteristics away from interfaces dominates, while at high resolution the error near the interfaces is dominant. Since discontinuities in the pressure gradients and the derivatives of the velocity field are present near the interfaces, the discretization schemes lose an order of accuracy there and the decrease of the order of convergence to $p=1$ is expected.

\begin{table}[ht]
\begin{center}
\begin{tabular}{ccc}
\hline
 Parameter & Value & Units \\
\hline
 $t^*={t}/{t_c}$ & $\displaystyle{\frac{t}{2.03}}$ & - \\
$E_{p,1}^f$ the potential energy in fluid 1 for $t \rightarrow  \infty$ & $1035$ & J \\
$E_{p,2}^f$ the potential energy in fluid 2 for $t \rightarrow  \infty$ & $3755$ & J \\
$\displaystyle{E_{k,1}^*=\frac{E_{k,1}}{(1/16) \rho_1 U_g^2 H^3}}$ & $\displaystyle{\frac{E_{k,1}}{3.41}}$ & - \\
$\displaystyle{E_{k,2}^*=\frac{E_{k,2}}{(1/16) \rho_2 U_g^2 H^3}}$ & $\displaystyle{\frac{E_{k,2}}{3.78}}$ & - \\
${{H^3}/{8}}$ the final volume of fluid 1 in the top part of the box & $0.125$ & $m^3$ \\
Maximum of enstrophy in fluid 1 (DyJeAT on a $2048^3$ grid) & $103$ & $m^{3}.s^{-2}$ \\
Maximum of enstrophy in fluid 2 (DyJeAT on a $2048^3$ grid) & $1049$ & $m^{3}.s^{-2}$ \\
\hline
\end{tabular}
\end{center}
\caption{Parameters used to define the dimensionless variables in case 2.}\label{tab4.2}
\end{table}

\begin{table}
\begin{center}
\begin{tabular}{cccc}
\hline
 $n$  & cells &$|E^*_p(n+1)-E^*_p(n)|$ & $p$  \\
\hline
  1 & $128^3$   & $0.0281$ & $1.6674$ \\
  2 & $256^3$   & $0.0088$ & $2.4120$ \\
  3 & $512^3$   & $0.0017$ & $0.9675$ \\
  4 & $1024^3$  & $0.0008$ & -      \\
  5 & $2048^3$  &   -    & -      \\
\hline
\end{tabular}
\end{center}
\caption{Case 2. Order of convergence $p$ of the dimensionless potential energy of fluid 1 at $t^*=2.63$.  \label{tab.conv}}
\end{table} 
After that time, the comparison of the three finest grids shows
that a well-characterized convergence no longer exists: the predicted kinetic energy varies
irregularly with the grid size, the difference between the $2048^3$ and $1024^3$ grids being often larger than that between the $1024^3$ and the  $512^3$ grids. This behavior starts at  $t^*=1.75$ and becomes pronounced beyond $t^*=5$. This can only indicate one thing: despite
the kinetic energy resulting mostly from large-scale motions,
the small scales influence strongly the evolution of the kinetic energy.
These small scales may for instance be involved in coalescence events, which themselves display the perforation of thin liquid sheets. Whether the sheets perforate and are destroyed or not has a strong influence on the large scales. Because of the observed convergence behavior of DyJeAT, its $2048^3$ kinetic energy can be taken as
  a reference, {with certainty until $t^*=1.75$, and as a reasonable estimate until $t^*=5$.

 An oscillating behavior is still present in the kinetic energy evolution, even though it
is less regular than in case 1.  In all
simulations, the kinetic energy decays as $t^{*-2}$ in fluid 2, as already found with case 1. In fluid
1, the exponential Stokes decay law (see Appendix A) is clearly more difficult to obtain, as large amplitude variations
are observed. Compared to case 1, the turbulent intensity in fluid 2 is larger, especially in the
bottom part of the box, and thus provides a strong forcing to the large blob of light fluid that stands
on top of it. This is why the oscillatory motion of fluid 1 is more complex and its
decay does not strictly follow the purely viscous Stokes law.

\begin{figure}[ht!]
\begin{center}
  \includegraphics[width=6cm]{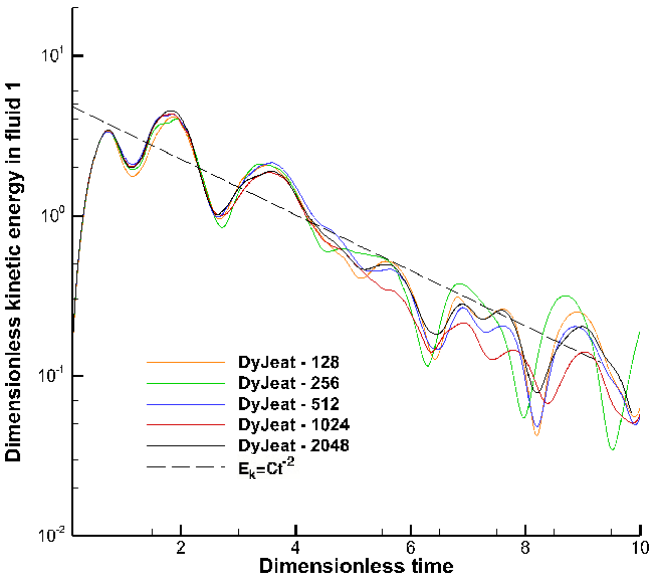}
  \includegraphics[width=6cm]{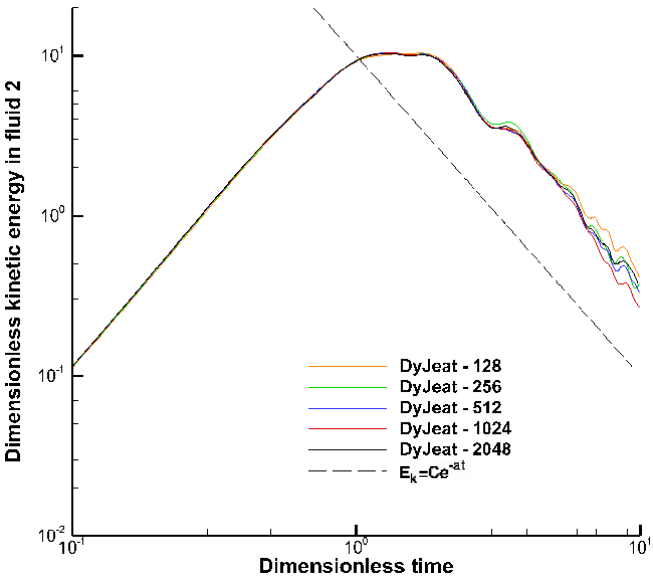}
  \end{center}
\caption {Grid convergence on kinetic energies for case 2 in log-linear (left) and log-log (right) coordinates.}
\label{fig4.10}
\end{figure}

\begin{figure}[ht!]
\begin{center}
  \includegraphics[width=6cm]{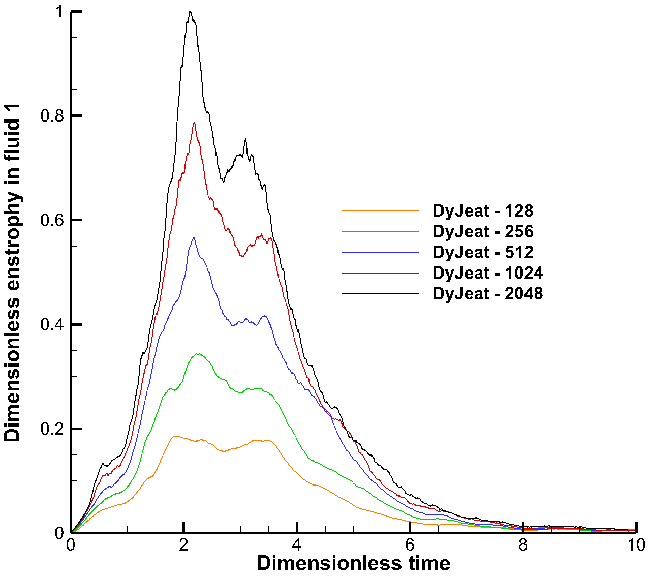}
  \includegraphics[width=6cm]{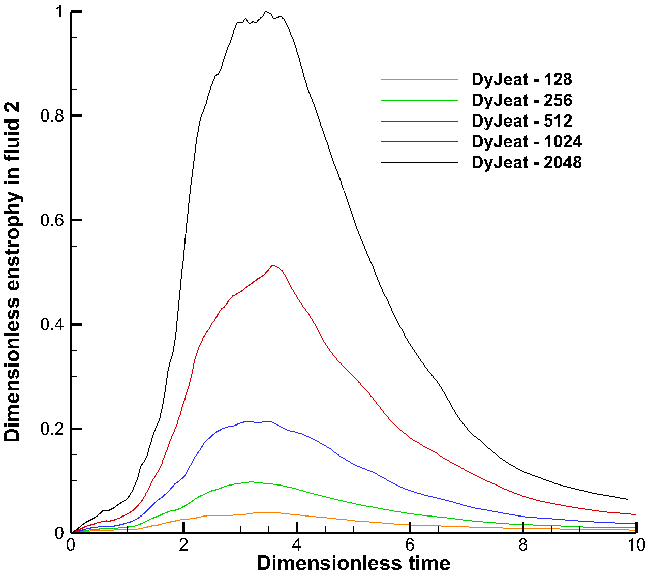}
  \end{center}
\caption {Convergence of enstrophy for case 2.}
\label{fig4.13}
\end{figure}

The time histories of enstrophy are plotted in Figure
\ref{fig4.13}.  The delay observed in the development of enstrophy in fluid 2 (compared to fluid 1) was not 
observed in case 1. This newly found delay is a consequence of the higher Reynolds number: vortical layers develop
quite quickly in the more viscous fluid 1, but a significantly longer time is required for shear
regions to develop in fluid 2, owing to the prevalence of inertial effects.

Again, the enstrophy
magnitude increases with the grid resolution. Compared to case 1, this tendency
is reinforced by the large number of break-up events which result in a large population of droplets of
fluid 1 that modulates the motion of fluid 2. Whatever the grid resolution, including $2048^3$,
convergence is not achieved and the finer the grid, the larger the enstrophy magnitude. As Figure
\ref{fig4.13} shows, the difference between the peak magnitudes obtained with two
successive resolutions increases as the grid is further refined. This is an indication that the highest
resolution considered here is still far from that required to achieve a true DNS. However,
it is observed that the width of the time interval in which the enstrophy
differs is decreasing with the refinement of the grid, testifying that grid convergence may be reached on
a finer mesh.

{A view of the interfacial structure is provided in Figure \ref{fig4.13bdyjeat}. Many lenticular (or ``flattened'' ) droplets are seen,
  an effect of the large rising speed that results in inertial effects frequently stronger than surface tension effects. This relative weakness of capillary effects occurs at fairly large scales, with a threshold that seems to be close to $0.05H$. \red{This length is close to the Hinze scale $\eta_H = 0.03H$ given in Table \ref{tab:nondim}, which was to be expected.} Smaller droplets are seen to remain nearly spherical. The interface rendered in Figure \ref{fig4.13bdyjeat} is also seen to be much smoother than in Figure \ref{fig1.1}, indicating that most of the droplets are well-resolved. We will return to the issue of the resolution of the smallest scales in Section \ref{sec.pdf}. }
\begin{figure}[ht!]
\begin{center}
  \includegraphics[width=8cm]{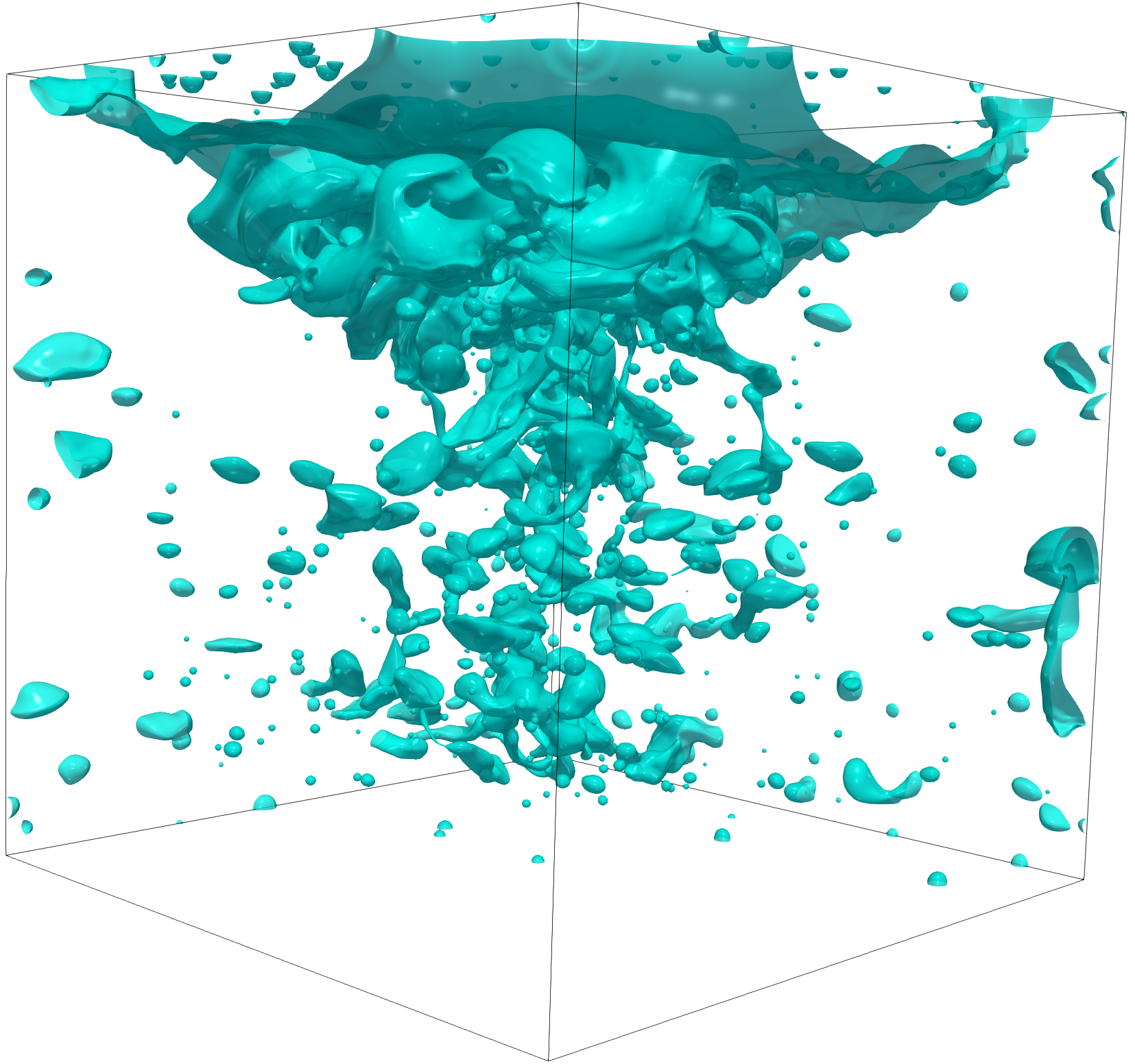}
  \end{center}
\caption {Topology of interfaces for case 2 at $t^*=3.5$ (near the enstrophy peak) in fluid 2 using DyJeAT on a $1024^3$ grid. The interface is much smoother than in Figure \ref{fig1.1}, indicating that most of the droplets are well-resolved.}
\label{fig4.13bdyjeat}
\end{figure}

%% Large interfacial scales are observed in the top part of the box, with at the same time an important population of small dispersed droplets of fluid 1. Consequently, high-shear regions are more numerous and difficult to capture than in case 1. The vorticity in fluid 2 is modulated by the presence of the light droplets of fluid 1 that first follow the motion of fluid 2 at short times, owing to inertia and viscous effects, and then rise under buoyancy effects. The finer the grid, the smaller the droplets captured by the computation. This feature has a direct influence on the intensity and density of vortical regions in the flow, since the latter are closely related to the phase distribution, \textit{i.e.} to the dispersion of fluid 1 in the present case. Therefore the behavior of the volume-averaged enstrophy depicted above is a direct consequence of this grid-dependent population of droplets.

\subsection{Droplet size distributions in case 2}
\label{sec.pdf}

We performed a study of the droplet size distributions in case 2 %at the time of the enstrophy peak ($t^* = 4$).
at various times. The corresponding results are summarized in Figure \ref{drop}.
\begin{figure}[ht!]
\begin{center}
  \includegraphics[width=8cm]{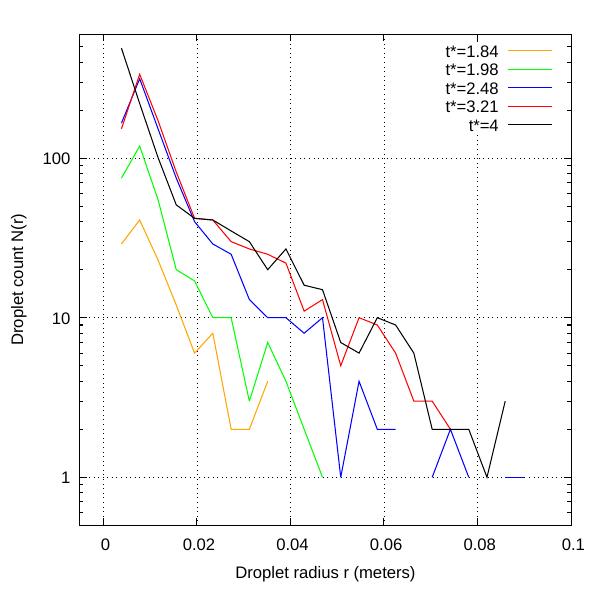}
  \end{center}
\caption {Droplet radius frequency $N_i$ at various dimensionless times obtained with DyJeAT at maximum resolution. The droplet radii are in meters, assuming the box size to be $H=1$\,m.}
\label{drop}
\end{figure}

{The droplet volumes $V$ are measured in the simulations
and a corresponding equivalent radius is obtained for each drop as $r_e = (3 V/(4 \pi))^{1/3}$. The equivalent radius may belong in any of a number of adjacent bins
$B_i$ defined as the interval $B_i = (r_i - \Delta_i/2, r_i + \Delta_i/2)$ where
$r_i$ and $\Delta_i$ are the bin center and width, respectively.}
We consider regularly spaced bins with $\Delta_i = 1/256$, $r_i = (i-1/2)/256$
and $B_i=((i-1)/256,i/256)$.
{Then the number of droplets whose equivalent radius lies in $B_i$ is noted $N_i$.
$N_i$ is approximately proportional to an asymptotic probability $f$ such that $f(r_i) = N_i/\Delta_i$; however this approximation is inappropriate for the first bin
  % in all cases % !!! SZ depending on the option
  because $\Delta_1/r_1$ is far from small. The function $f(r_i)$ defines the Probability Distribution Function (PDF) of the droplet sizes. 
  In Figures \ref{drop} and \ref{fig4.13b}, we plot $N_i$ versus $r_i$ at various times or various resolutions. }

\begin{figure}[ht!]
\begin{center}
  \includegraphics[width=8cm]{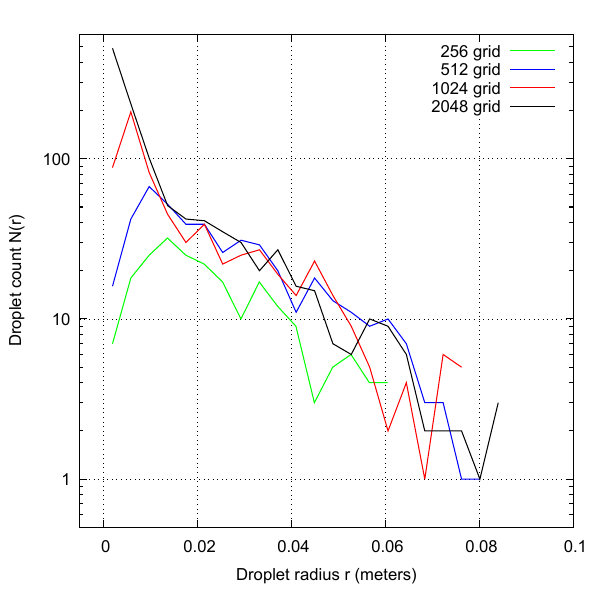}
  \end{center}
\caption {Droplet radius frequency $N_i$ obtained with DyJeAT at
  $t^*=4$. The droplet radii are in meters, assuming the box size to be $H=1$\,m.}
\label{fig4.13b}
\end{figure}
    {The observation of these distributions leads to several remarks.
      First, as time progresses (see Figure \ref{drop}), the number of droplets increases, but the number
      of large drops increases faster than the number of small droplets. Beyond a radius of 0.02\,m, the distribution roughly follows an exponential law with an average droplet size $\langle r \rangle$ that increases in time. }\\
{
In Figure \ref{fig4.13b}, a convergence study of the droplet size distribution is performed at time $t^* = 4$. 
First, it is seen that the $256^3$ grid result does not match the results obtained on finer grids. Instead, droplet populations
obtained with that grid are systematically
less numerous than with the other grids, indicating that the total mass
of droplets obtained with that grid is smaller. This can be easily explained by the fact that
poorly resolved droplets ``evaporate'' in a Level Set method such as that used in DyJeAT. 
Second,} the $512^3$, $1024^3$ and $2048^3$ grid results almost superpose in
a region of variable size, going from a minimum radius $r_m(N)$ to a maximum
radius $r_M(N)$, with $r_m$ and $r_M$
both depending on the grid resolution $N$.
To be specific,
the minimum radius of agreement decreases with increasing $N$ until approximately
$r_m(2048) \simeq 0.005$\,m
which indicates a range of convergence extending to very small droplet radii
for the $2048^3$ grid. The upper boundary $r_M(N)$ of the range of convergence is harder
to determine, because large radii are affected by statistical noise. However,
it is seen that few  additional very large droplets ($r > 0.07$\,m) are identified as the resolution increases from
$256^3$ to $2048^3$. Moreover, grids $1024^3$ and $2048^3$ are in agreement over a wide range
with $r_M(2048) \simeq 0.05$\,m. Finally, we note that the behavior at very small radii differs between
  the  $2048^3$ grid and the other grids: while the latter have a cutoff (a fall in droplet counts $N_1$ and/or $N_2$) at small sizes, the  $2048^3$
  grid does not.
This is due to the fact that in the $2048^3$ case, the fall in droplet count
can only be seen if instead of the peculiar bin $B_1$ defined above, one subdivides the interval $(0,1/256)$ into several smaller bins.
We have actually done that and did observe the cutoff even for the  $2048^3$ case. 

\begin{figure}[ht!]
\begin{center}
  \includegraphics[width=8cm]{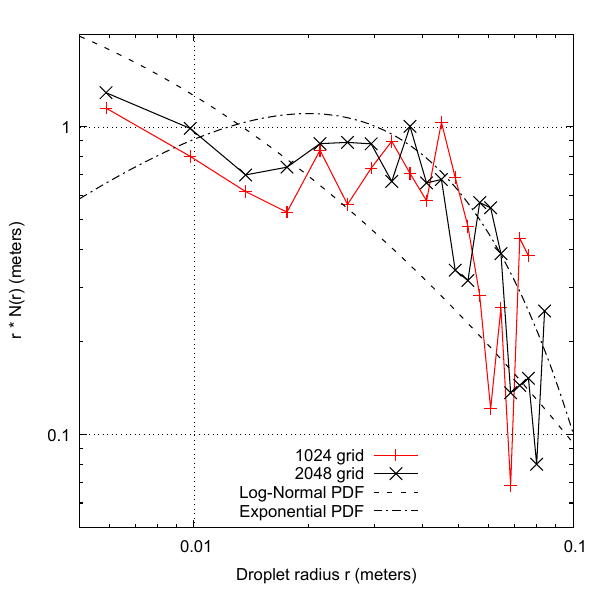}
  \end{center}
\caption {Droplet radius frequencies (PDF)
  in logarithmic coordinates, with $r f(r)$ versus $r$. A log-normal distribution and an exponential
  one are shown to guide the eye.}
\label{fig4.13ln}
\end{figure}

{Similar effects of the grid refinement,
including cutoffs or drops in frequency,
have been reported for atomization of liquid jets, especially with Level Set methods
\cite{herrmann2011simulating} and less markedly for VOF approaches \cite{ling17}.}
This indicates that $512^3$ grids are a minimum requirement to reach an approximately
accurate droplet count at {\em any} scale, and that $256^3$ simulations are poorly
resolved in the whole range of scales.
On the other hand, the absence of convergence of the enstrophy does not imply a non-convergence of the droplet sizes in the intermediate range
$r_m(N) < r < r_M(N)$.
Another interesting fact is that the large scales are influenced by how the small scales
are computed: more large droplets are seen with finer grids,
probably because a higher grid resolution implies less breakup of thin layers or filaments,
thus reducing the rate at which large structures break into smaller ones.
In order to better understand the PDF, we plot the computed frequencies for the two most refined
grids as a graph of $\ln (r_i N_i)$ versus $\ln r_i$ in
Figure \ref{fig4.13ln}.
The first bin $B_1$ was excluded due to its particular character, and only results for the most refined
grids $1024^3$ and $2048^3$ are shown. 
Using these coordinates has the advantage that the log-normal PDF appears as a parabola.
This specific PDF is defined as
\begin{equation}
  f(r) = \frac A {r} \exp \left[ - \frac{(\ln r  - \ln \hat \mu)^2 }{2 \hat \sigma^2} \right]\,,
\end{equation}
where $A$ is a normalization constant, $\ln \hat \mu$ is the logarithmic average and $\hat \sigma$ is the logarithmic
standard deviation. A parabolic curve corresponding to such a log-normal distribution is shown on Figure  \ref{fig4.13ln}
to aid the interpretation. This parabola is built with $\hat \mu = 10^{-3}$\,m, $\hat \sigma=1.7$ and $A=3$.
While the value of $\hat \mu$ is very uncertain, it is clear that extrapolating the trend of the distribution
leads to expect a larger value of $r f(r)$ at smaller $r$. These larger values of $r f(r)$ could be revealed in simulations
with yet more refined grids than
the current ones. This strongly reinforces the expectation that many smaller droplets would be seen in a true DNS.
However, these droplets would contribute little to the interfacial area. Indeed the interfacial area scales as
\begin{equation}
  \Sigma_S  = \int_{0}^{\infty} 4 \pi r^2 f(r)\, {\rm d}r \,,
\end{equation}
which for a log-normal $f(r)$ or a simpler $f(r) \sim 1/r$ converges at the $r=0$ bound.
Note that we took the assumption of spherical droplets to estimate $ \Sigma_S$. The latter assumption is likely to be a good
approximation for the smallest droplets. 
Thus, the absence of such droplets in the integral does not affect much the interfacial area.
 
{A very simple Pareto distribution of the form  $f(r) \sim A/r$ would be a horizontal line in the variables
of Figure \ref{fig4.13ln}. That approximation is in fact a good fit in the 
intermediate range of droplet sizes with $A \simeq 1$, and is reminiscent of distributions seen in other contexts
\cite{BALACHANDAR2020103439,pairetti2021numerical}. As a third alternative, an exponential distribution $f(r) = B \exp(-r/r_m)$
is also plotted on Figure \ref{fig4.13ln} with $r_m=0.02$\,m. The exponential distribution provides
a better fit for the largest $r$. With this distribution, the interfacial area integral
converges for $\Delta x \ll r_m = 0.02\,$\,m and thus for $256^3$ grids. 

{Finally, we note that none of the distributions yields a very good fit throughout the entire range of scales. It is possible that the PDF is in fact bimodal, with
  different ``modes'' corresponding to different physical mechanisms being superimposed at various scales.

%%================================================================================

% Part 2: Results using all the codes

%%================================================================================

\section{Results of the benchmark using all codes at moderate resolution}

The teams involved in the present benchmark ran their own code each.
%These CFD tools are running with different speed-up parallel properties and CPU times.
Throughout, $256^3$ grids were used.

%----------------

% CASE 1 all codes

%----------------
\subsection{Case 1 results for all codes}

{We first compare the performance of all codes in terms of mass conservation in Figure \ref{mass-c1}. DyJeAT is also
run at $256^3$ resolution for this purpose. In the case of DyJeAT, the issue of mass conservation is quite specific, since the mass decrease or increase is computed at each time step before the level-set function is shifted
to bring the mass deviation back to zero. \red{In doing so, mass may be transferred in a non-local manner. Thus the interface is rearranged locally,
making for instance small droplets disappear. }
What is 
plotted in Figure \ref{mass-c1} in the case of DyJeAT is the cumulative absolute value of the deviation before the level-set shift.
All codes have \red{fair} mass conservation properties with the worse volume error always
smaller than \red{$1.6\%$ of the volume of light fluid. This is a nevertheless quite significant error that should be reflected in energy conservation, since conservation of the total mechanical energy is closely related to volume and mass conservation. Here, since almost all of fluid 1 goes to the top, the final potential energy of the system, equal to the final total mechanical energy minus the surface energy, is determined to a very good approximation by the final mass of fluid 1 in the top region. Thus, any error in the final volume (hence the mass) of fluid 1 in the top region should result in an error on the final mechanical energy. \\
Figure \ref{em-allcodes}  shows that all codes predict a final mechanical energy close to the reference value. More precisely, normalizing differences by the total energy variation from $t=0$ to $t\rightarrow\infty$, it is found that the departure from the reference value is less than $1.4\%$ for all codes, which is consistent with the error found on mass conservation. More specifically,} it is seen 
  %However as we saw above, conservation of the total mechanical energy is closely related to volume and mass conservation. It is shown here for all the codes in Figure \ref{em-allcodes}. Indeed the final potential energy, equal to the final total mechanical energy minus surface energy, is determined to a very good approximation by the final volume of fluid 1 in the top region. Thus if volume is conserved, and since almost of all of fluid 1 goes to the top, mechanical energy will also be conserved. Conversely the observed energy conservation implies mass conservation. }
in Figure \ref{em-allcodes} that
  Archer and Jadim agree with the reference within the thickness of the line, while Thetis and Gerris perform somewhat worse,
  with Thetis even below the theoretical asymptotic value. This ranking of the codes may be partially inferred from Figure \ref{mass-c1} which confirms the good mass conservation behavior of Archer and Jadim
  but does not show a significantly worse behavior for Thetis. Indeed there are other aspects that can affect the
  final potential energy: droplets of light fluid that remain attached to the bottom wall by capillary forces
  increase the final potential energy, as do droplets of heavy fluid attached to the \red{top}.
  This effect is counterbalanced by mass conservation: an excess (resp. deficit) of light fluid at the end of the simulation, near the top, decreases (resp. increases) potential energy. It is seen that the mass of light fluid increases
  with Thetis, which agrees with the undershoot seen in the corresponding energy. Finally, DyJeAT with a $256^3$ resolution is the only code that is
  slightly above the  theoretical asymptotic value. This is probably related to the specific treatment of mass conservation described above. 
}

\begin{figure}[ht!]
\begin{center}
  \includegraphics[width=9cm]{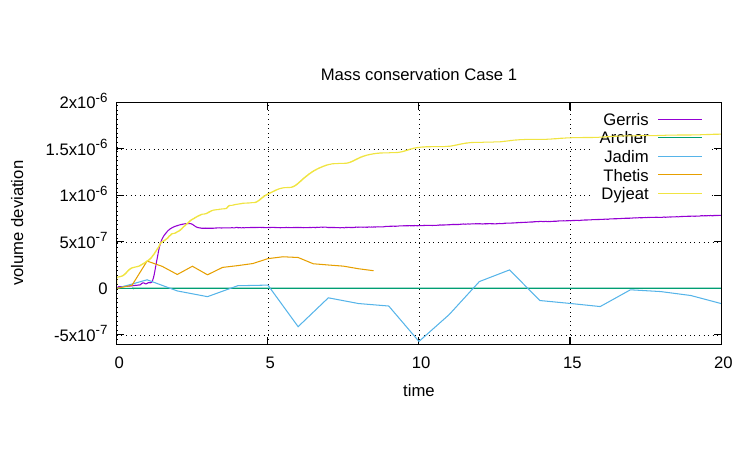}
  \end{center}
\caption {Mass conservation for case 1. The dimensional volume deviation of fluid 1 (light fluid) is plotted as a function of the dimensionless time.}
\label{mass-c1}
\end{figure}

\begin{figure}[ht!]
\begin{center}
  \includegraphics[width=6cm]{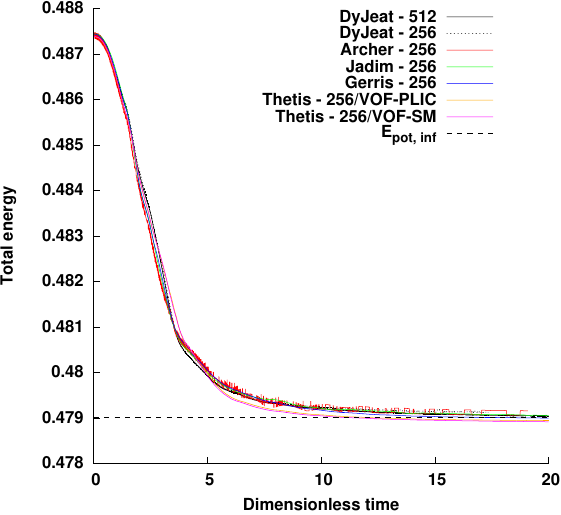}
  \end{center}
\caption {Evolution of the dimensional total mechanical energy in case 1 (without the surface energy), $E_m$, for all codes as a function of the dimensionless time.
The horizontal line is the expected value $E^f_{p}$. }
\label{em-allcodes}
\end{figure}

Three macroscopic quantities, namely the kinetic and potential energies and the volume of fluid 1 ``in the top part''
were computed with the four codes other than DyJeAT.
\begin{figure}[ht!]
\begin{center}
  \includegraphics[width=6cm]{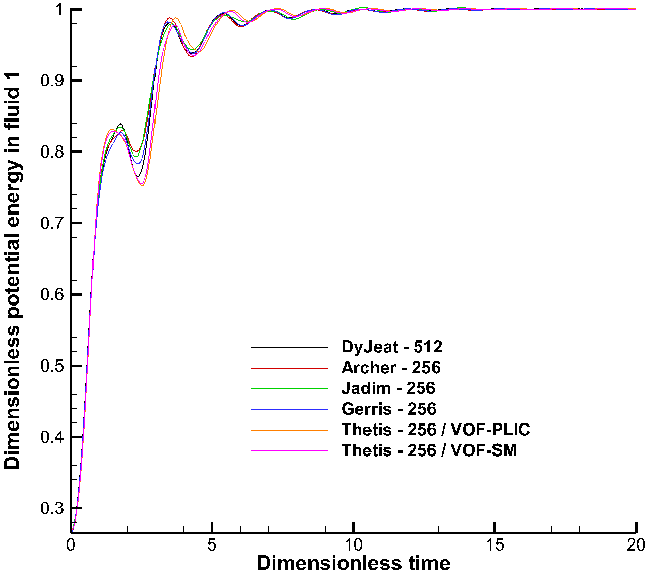}
  \includegraphics[width=6cm]{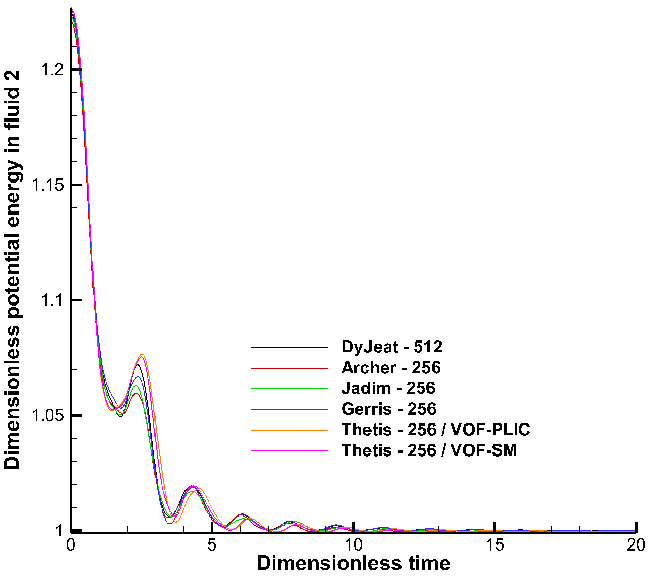}
  \end{center}
\caption {Potential energies for case 1.}
\label{fig4.1}
\end{figure}
\begin{figure}[ht!]
\begin{center}
  \includegraphics[width=6cm]{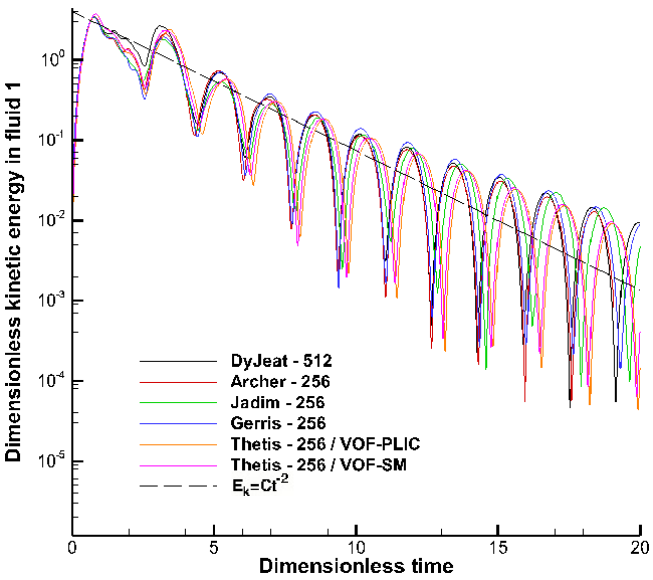}
  \includegraphics[width=6cm]{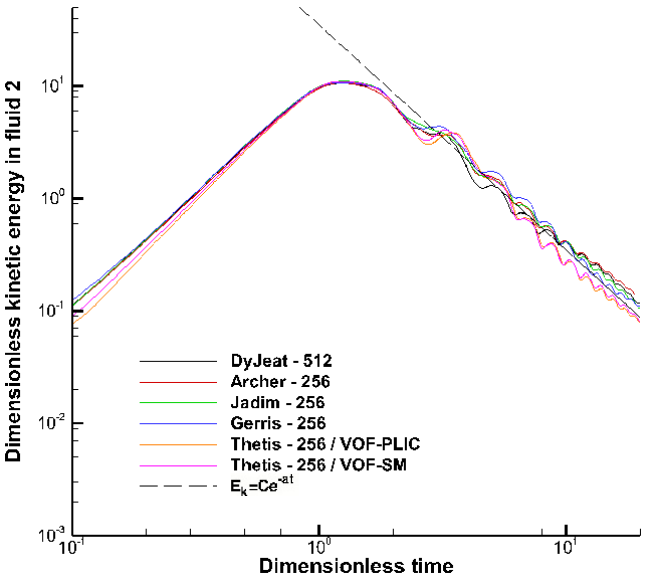}
  \end{center}
\caption {Kinetic energies for case 1 in log-linear (left) and log-log (right) coordinates.}
\label{fig4.2}
\end{figure}
\begin{figure}[ht!]
\begin{center}
  \includegraphics[width=6cm]{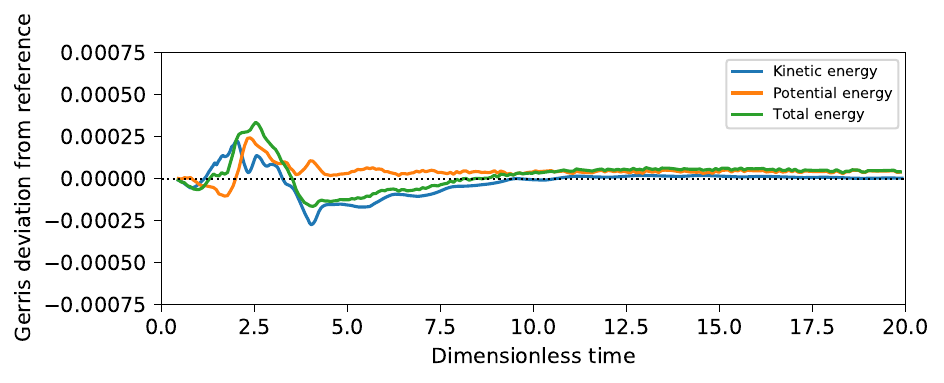}
  \includegraphics[width=6cm]{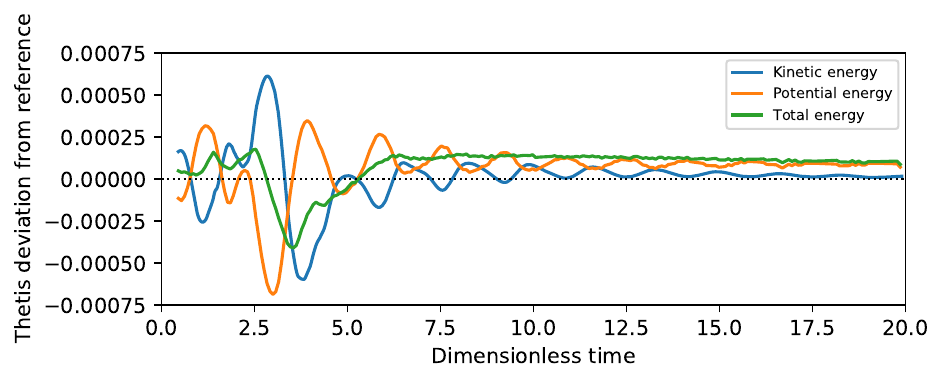}
  \includegraphics[width=6cm]{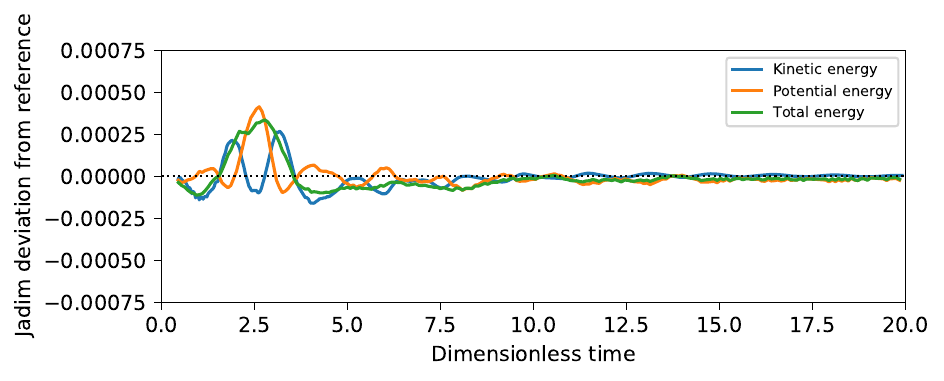}
  \includegraphics[width=6cm]{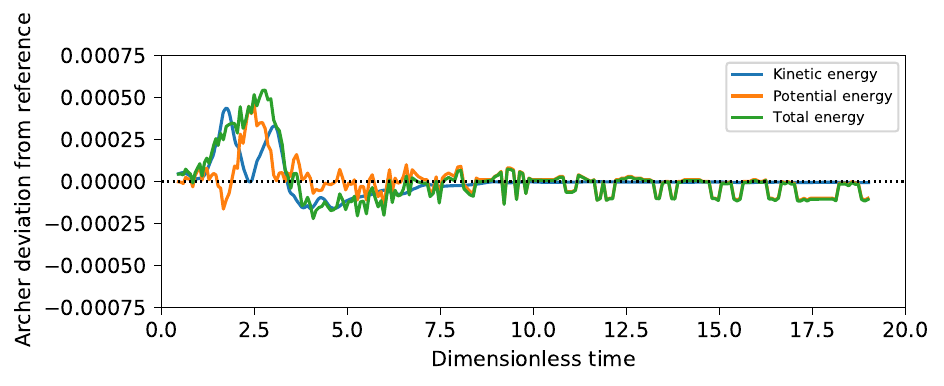}
  \includegraphics[width=6cm]{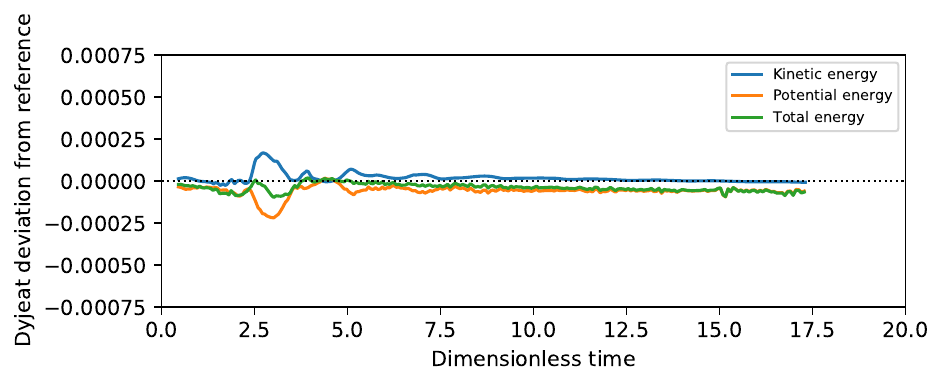}
  \end{center}
\caption {Deviation of the energies from the reference in case 1. DyJeAT $256^3$ is compared to the DyJeAT $512^3$ reference.}
\label{newplotscase1}
\end{figure}
%% \begin{figure}[ht!]
%% \begin{center}
%%   \includegraphics[width=6cm]{Cas3_EVolFluid1.png}
%%   \end{center}
%% \caption {Evolution of the volume ratio of fluid 1 in the top part of the box in case 1 (all codes).}
%% \label{fig4.4all}
%% \end{figure}
{The evolution of the potential and
kinetic energies is plotted in Figures \ref{fig4.1} and \ref{fig4.2},
respectively, together with the DyJeAT reference solution. The
results provided by all five codes on various grids are in approximate agreement with each other and
with the reference solution. Around time $t^* = 3$, differences are seen between
all codes during the first oscillation arc of the potential energy. At later times, the
oscillations of all codes are superposed, except for the two versions of Thetis, and to a lesser
degree the Jadim code. The behavior of the kinetic energy is more complex as differences
between the behavior in fluids 1 and 2 are noticed. However, at large times the kinetic energy predicted by the
Archer code for both fluids is the closest to the reference solution. This may be due to the non-converged
character of the DyJeAT reference solution for {$t^* > 2.5$}, which would make the agreement with Archer easier
since both codes are at least partially based on the Level Set approach. In other words, Archer being the code
most similar to DyJeAT is the most likely to agree with the reference. 
Figure \ref{fig4.1} indicates that phase separation is achieved after a dimensionless time $t^*\approx15$. The dimensionless frequency of the waves observed at the surface of the light fluid is about $0.6$. }
Again, it is observed that all codes predict evolutions close to each other and to the
reference solution. \vspace{1mm}\\
{To better view the various predictions of kinetic and potential energies by the different codes in another perspective,
we plot in Figure \ref{newplotscase1} the deviation of the energies of all codes with respect to the DyJeAT reference.
This deviation is shown for the kinetic energy, the potential energy, and the sum of these two components (total mechanical energy minus the surface energy). Interestingly, for some of the codes (especially Thetis) \red{the differences are oscillatory
while for others there is a less coherent drift.}} {Note also that the energy deviations shown in   Figure \ref{newplotscase1} are of $\Order(10^{-3})$ and thus much smaller than the total potential energy which is of $\Order (0.1)$ (This value of the potential energy in Joules can be deduced from the characteristic energies
used for the normalization defined in Table \ref{tab4.1}.) }\\
\color{red} The code-to-code comparison provided by Figures \ref{newplotscase1} indicates some phase shifts and various damping rates at large times.
In this limit, Thetis is the furthest from the DyJeAT reference, followed by Gerris, while Jadim is very close to that reference. Archer has fluctuations that make the energy at large times slightly unsteady.
The codes exhibit the largest difference with the reference at time $t^*=2.5$, close to  the kinetic energy peak (see Figure \ref{fig4.2}). These differences, while all of the same order of magnitude are clearly the largest for Thetis followed by Archer, then Jadim and Gerris, and are the smallest for DyJeAT $256^3$. 
A possible explanation for these code-to-code differences could be the quality of mass conservation in each of them,
since it directly influences the amount of fluid in the top part. Another explanation would be the accuracy of
energy conservation since it drives the observed oscillations. A third explanation is
the accuracy of the computation of the surface tension force, since it can remove or add energy proportional to (i)  the
interfacial area and (ii) the ``spurious currents'' generated by small imbalances with the other terms in the momentum equation (see \cite{tryggvason11}).  \color{black}

%%The evolution of the volume ratio of fluid 1 in the top part of the box is shown for all the codes in Figure \ref{fig4.4all}. 

\begin{figure}[ht!]
\begin{center}
  \includegraphics[width=6cm]{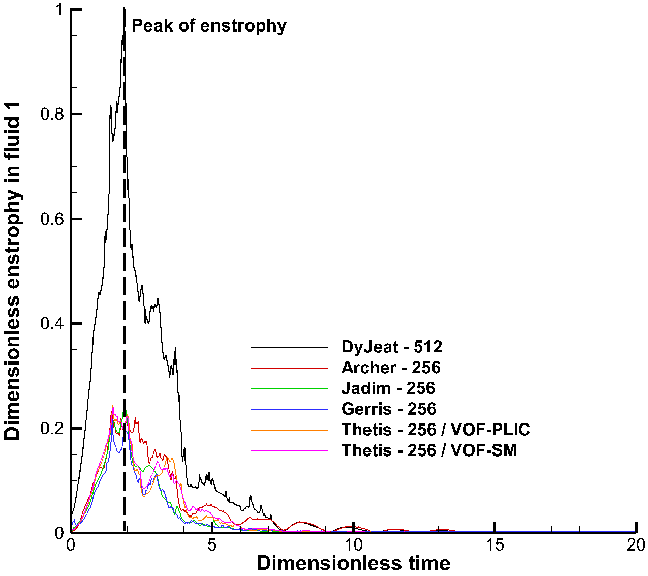}
  \includegraphics[width=6cm]{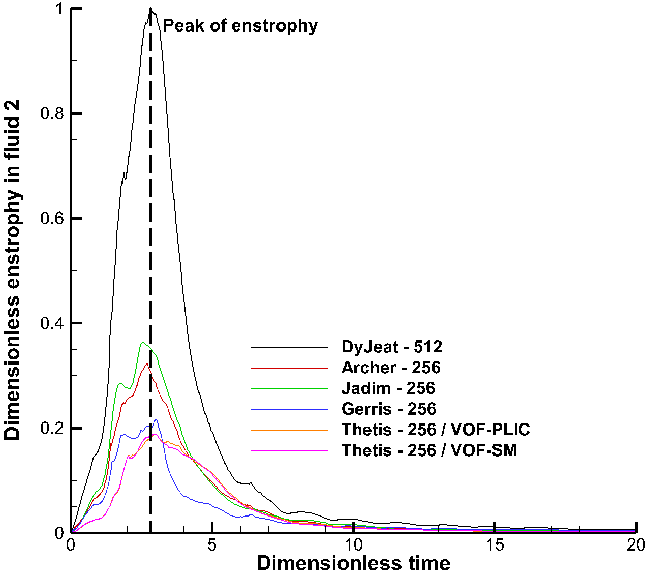}
  \end{center}
\caption {Enstrophy for case 1.}
\label{fig4.5}
\end{figure}

\begin{figure}[ht!]
\begin{center}
  \includegraphics[width=6cm]{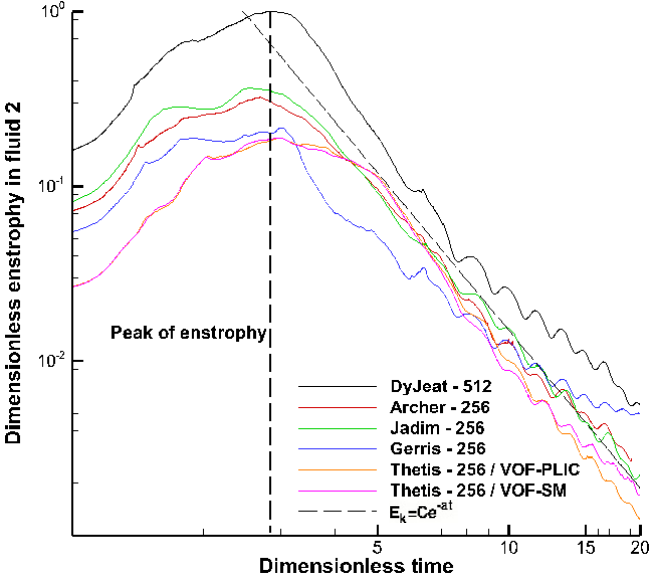}
  \end{center}
\caption {Enstrophy for case 1 in log-log coordinates.}
\label{fig4.5b}
\end{figure}

%\begin{figure}[ht!]
%\begin{center}
%  \includegraphics[width=8cm]{Cas3_AirInt1.png}
%  \includegraphics[width=8cm]{Cas3_AirInt2.png}
%  \end{center}
%\caption {Interfacial area for case 1.}
%\label{fig4.7}
%\end{figure}

Figures \ref{fig4.5} and \ref{fig4.5b} display the instantaneous values of the enstrophy in both fluids. Although all codes were shown to agree well on other macroscopic quantities, the magnitude of the enstrophy peak differs dramatically for each of them. The location of this peak is found to occur at $t^*\approx 2$ (resp. $3$) for all codes in fluid 1 (resp. fluid 2). However, differences in magnitude up to $80\%$ are observed, depending on the grid and the code.  Despite the large discrepancies noticed between all codes, it is interesting to observe that they all provide a $t^{*-3}$ temporal decay law which may be shown to correspond to a turbulent scaling (see Appendix \ref{AppendixA}).\\

%----------------

% CASE 2 all codes

%----------------
\subsection{Case 2 results for all codes}

{We first compare all codes in terms of mass conservation in Figure \ref{mass-c2}. DyJeAT is also
  run at $256^3$ resolution for this purpose. The same remark as above applies to the technique involved for mass conservation in DyJeAT.
  The codes have very dispersed mass conservation properties. 
As above, conservation of the total mechanical energy approximately deduced from the analysis of Figure \ref{em2-allcodes}
indicates how each code conserves volume,
since the final potential energy is related to the final volume of fluid 1 in the top region.
It is seen in Figure \ref{em2-allcodes} that
  Archer and Gerris agree with the reference within the thickness of the line, while Thetis and Jadim perform worse,
  with Thetis again  below the theoretical asymptotic value. Unlike case 1, this ranking of the codes can also be deduced from the
  mass conservation error plotted in Figure \ref{mass-c2}, where Gerris and Archer perform much better than the other codes.
  In this case, both Gerris and DyJeAT $256^3$ end up above the DyJeAT reference value at late times.
 \red{The reason why mass conservation deteriorates in Jadim beyond $t^*=5$ is that the  local mass error control strategy described in Section \ref{jadimsection} was not designed to handle topological changes, and presumably fails during coalescence and breakup events.  Such events being much more frequent in case 2 than in case 1, departures from exact mass conservation are much larger in the former case.}  
}

\begin{figure}[ht!]
\begin{center}
  \includegraphics[width=9cm]{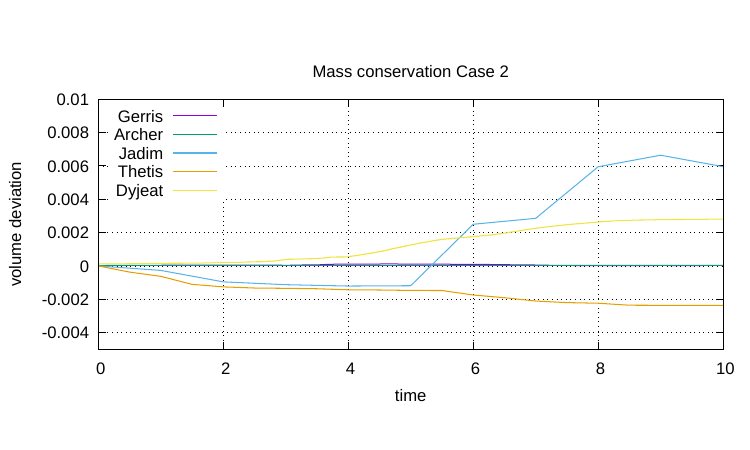}
  \end{center}
\caption {Mass conservation for case 2.}
\label{mass-c2}
\end{figure}

\begin{figure}[ht!]
\begin{center}
  \includegraphics[width=6cm]{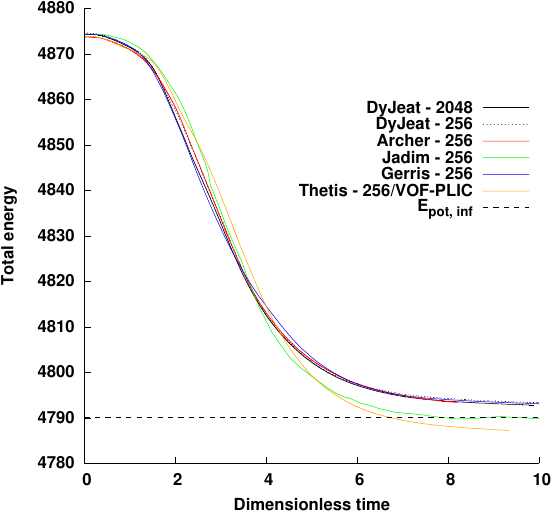}
  \end{center}
\caption {Evolution of the dimensional total mechanical energy in case 2 (without the surface energy), $E_m$, as a function of dimensionless time for all the codes. 
The horizontal line is the expected value $E^f_{p}$.}
\label{em2-allcodes}
\end{figure}

Similar to the observations reported for case 1, the results provided by the five codes are in approximate
agreement as far as the time evolution of the potential and kinetic energies is concerned, even if
larger discrepancies are observed compared to case 1. These evolutions are displayed in Figures
\ref{fig4.8} and \ref{fig4.9}.
{The behaviors of the codes are diverse. For example, the Gerris code performs well for the
kinetic energy of fluid 2.
However, it predicts a significantly different oscillatory dynamics for the kinetic energy of fluid 1
beyond $t^*=1.5$.}
In fact, for fluid 1, all codes
are close to the reference, except Gerris, until time $t^*=2.5$. Then they all diverge from
the reference. For fluid 2, Gerris and Thetis behave better than the other codes within the time interval $2 < t^* < 3$.

{Similar to case 1, to view the different predictions of kinetic and potential energies by the different codes in another perspective,
we plot the deviation of the energies predicted by all codes with respect to the DyJeAT reference in Figure \ref{newenplots}.
This deviation is shown for the kinetic energy, the potential energy, and the sum of these two components (total mechanical energy minus the surface energy). For some of the codes (especially Gerris), 
 the sum of the kinetic and potential energies remains close to the reference at early times, with both components
deviating in an oscillatory manner and in phase opposition. In the case of Gerris, the plot also includes the deviation
of the surface energy from the ``reference'' DyJeAT case. This deviation mainly serves to show that the surface energy fluctuations are of the same order as the deviations in kinetic and potential energies. It does not seem that the performance of the codes in terms of prediction of the
kinetic and potential energies is related to their performance in the accounting for surface tension forces, since,
at least in the case of Gerris, the deviations in these various energies are not correlated.} 
{For some other codes, the deviation from the reference is much less oscillatory, indicating that the error is dispersed in a non-oscillatory manner. Finally, in some cases the total mechanical energy (minus the surface energy) is not conserved, indicating a possible dissipative error or a mass conservation error.  \red{In the long time limit, Thetis and to a lesser extent Jadim have energies that deviate from the reference, while Gerris and Archer are very close to it. This behavior is consistent with the mass conservation of these codes. DyJeAT $256^3$ is close to the reference (obtained with DyJeAT itself). Unlike case 1, there is no clear maximum of the error around $t^* =2$ or 3 for all the codes, although maxima are found around $t^* = 2$ for some of them and either the total, potential or kinetic energies. These ``maximum errors" are of the same order of magnitude, between 2.5 and 5 Joules. 
Note also that these energy deviations need to be compared with the total kinetic
energy which is of the order of 40 Joules (see Table \ref{tab4.2} and Figure \ref{fig4.9}).} }

\begin{figure}[ht!]
\begin{center}
  \includegraphics[width=6cm]{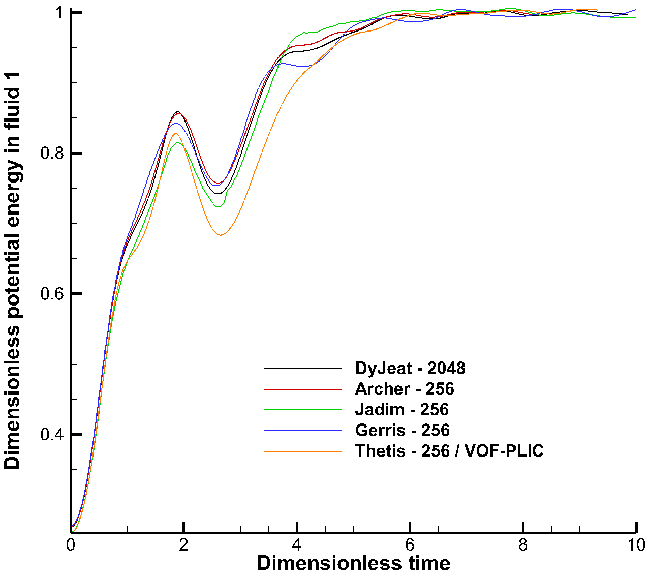}
  \includegraphics[width=6cm]{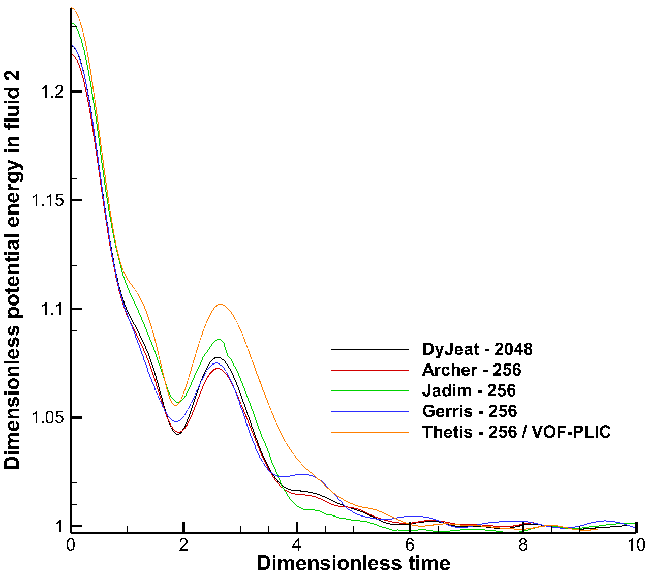}
  \end{center}
\caption {Potential energies for case 2.}
\label{fig4.8}
\end{figure}

\begin{figure}[ht!]
\begin{center}
  \includegraphics[width=6cm]{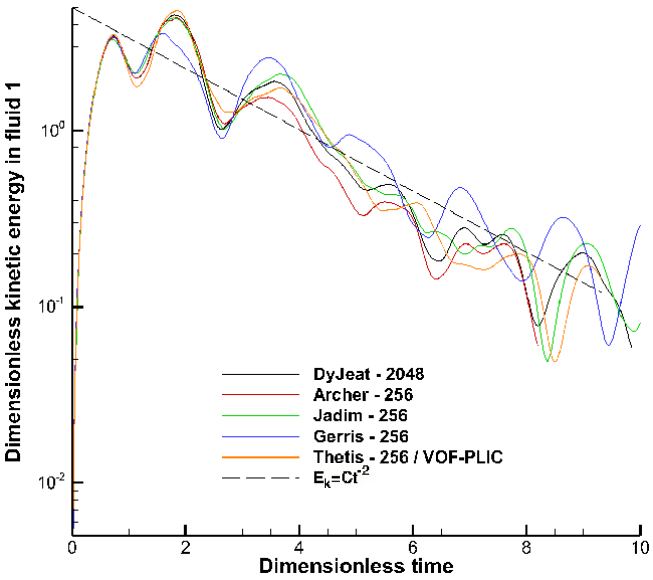}
  \includegraphics[width=6cm]{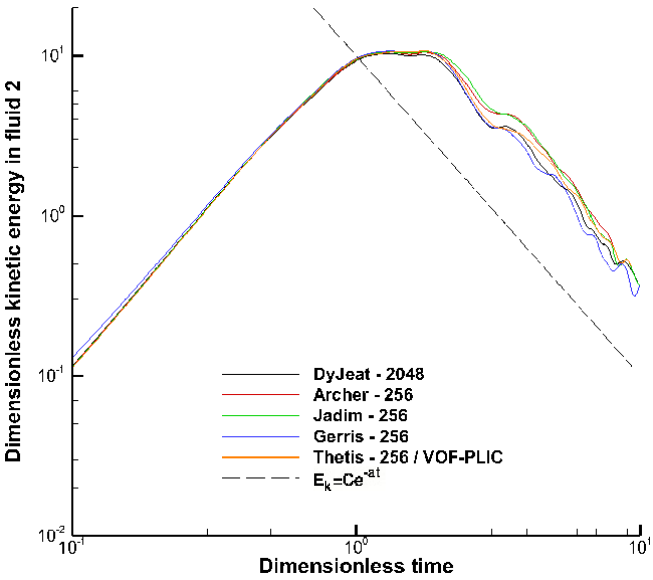}
  \end{center}
\caption {Kinetic energies for fluid 1 in log-linear coordinates (left) and fluid 2 in log-log coordinates (right) (case 2).}
\label{fig4.9}
\end{figure}

\begin{figure}[ht!]
\begin{center}
  \includegraphics[width=6cm]{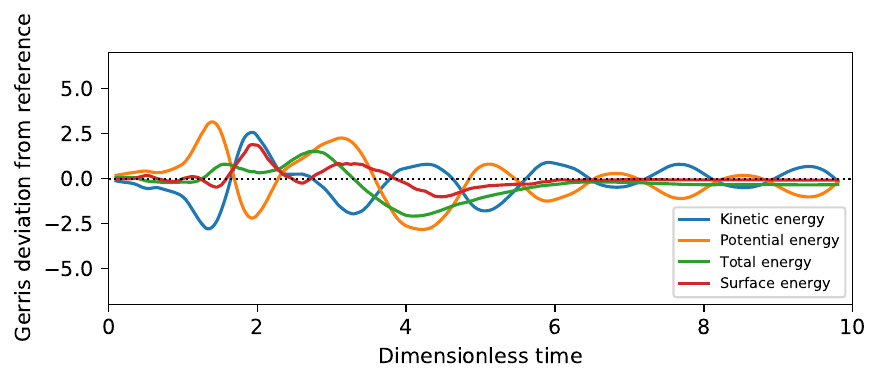}
  \includegraphics[width=6cm]{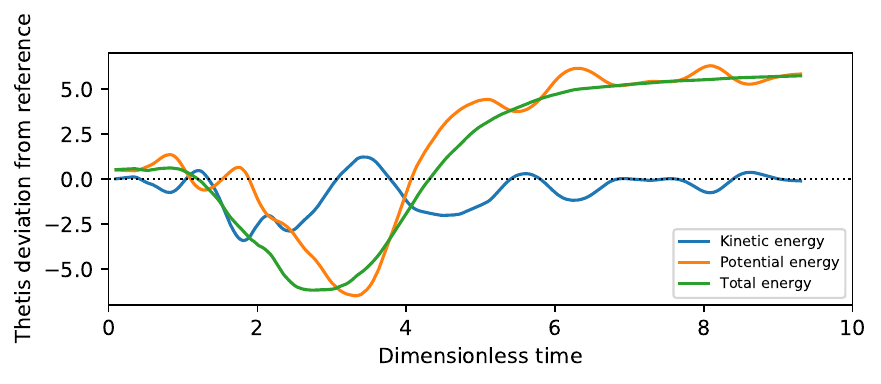}
  \includegraphics[width=6cm]{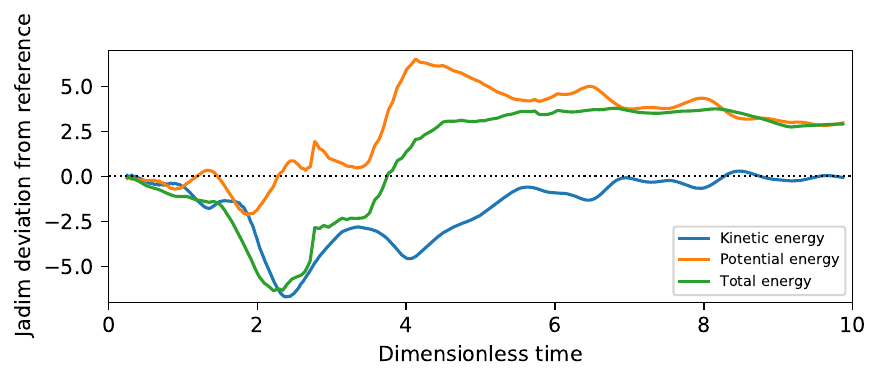}
  \includegraphics[width=6cm]{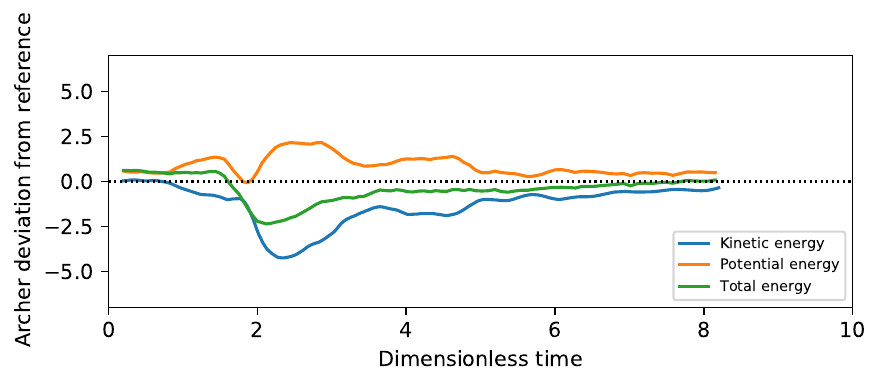}
  \includegraphics[width=6cm]{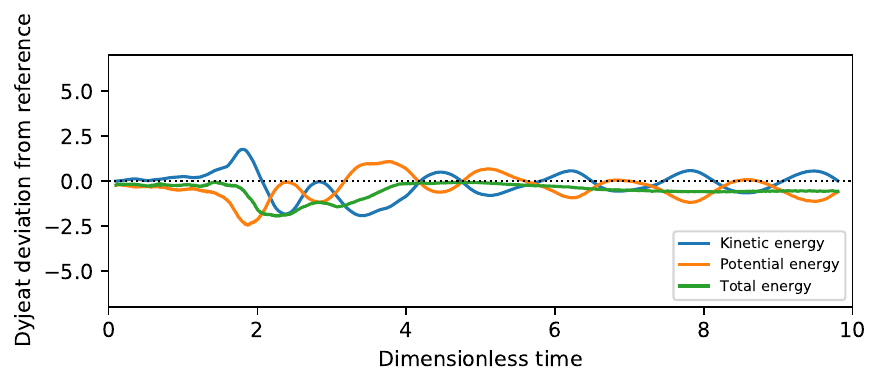}
  \end{center}
\caption {Deviation of the energies from the reference in case 2. Results shown for Gerris also include the deviation
  of the surface energy from the ``reference'' DyJeAT case. DyJeAT $256^3$ is compared to the DyJeAT $2048^3$ reference.}
\label{newenplots}
\end{figure}

%\begin{figure}[ht!]
%\begin{center}
%  \includegraphics[width=8cm]{Cas8_EVolFluid1.png}
%  \includegraphics[width=8cm]{Cas8_EVolFluid1b.png}
%  \end{center}
%\caption {Volume ratio of fluid 1 in the top part of the cavity during time for case 2.}
%\label{fig4.11}
%\end{figure}

\begin{figure}[ht!]
\begin{center}
  \includegraphics[width=6cm]{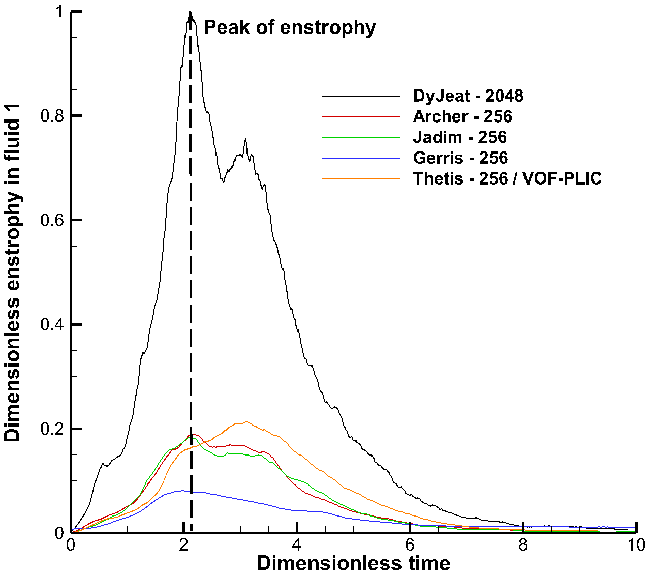}
  \includegraphics[width=6cm]{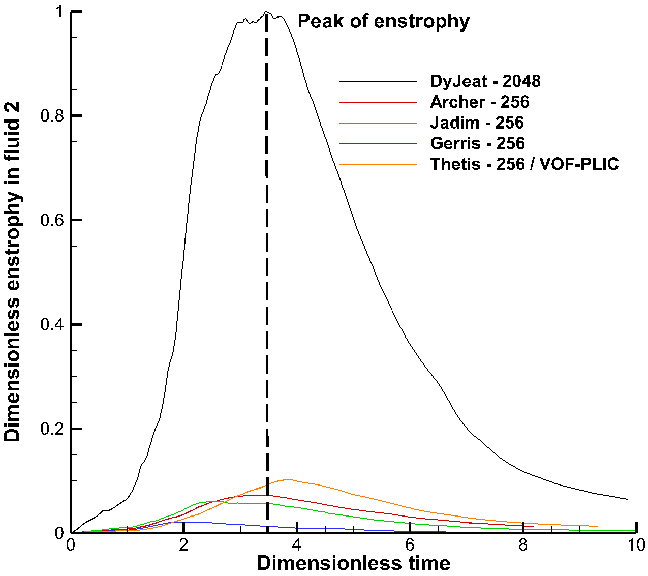}
  \end{center}
\caption {Enstrophy for case 2.}
\label{fig4.12}
\end{figure}

\begin{figure}[ht!]
\begin{center}
  \includegraphics[width=6cm]{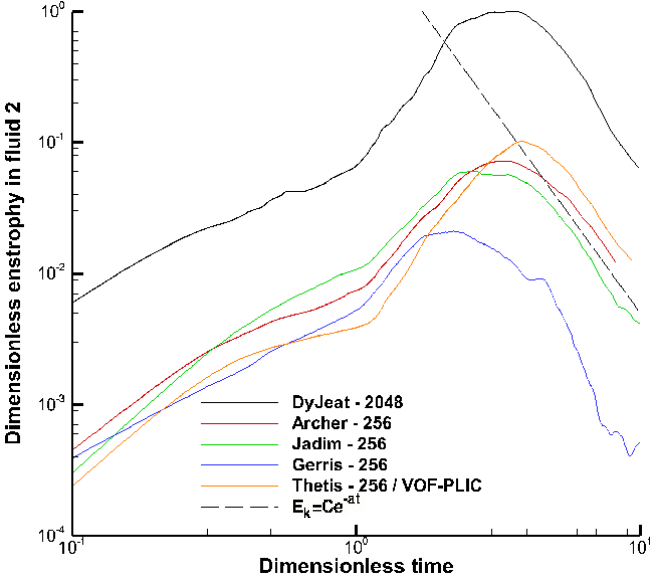}
  \end{center}
\caption {Enstrophy for case 2 in log-log coordinates.}
\label{fig4.12b}
\end{figure}

The time histories of enstrophy are plotted in Figures \ref{fig4.12}, \ref{fig4.12b} and
\ref{fig4.13}. Again, all codes find the peak value nearly at the same time, namely $t^*\approx2.5$ in
fluid 1 and $t^* \approx 4$ in fluid 2, although the code-to-code differences are significantly larger
than in case 1. Again, the enstrophy
magnitude increases with grid resolution and is code-dependent. \\
%\subsection{PDF of droplet sizes for all codes}
{The uncertainty surrounding the
  $256^3$ grid results for the PDF of droplets sizes leads us not to discuss the results provided by the other codes for this PDF, since these codes were not run at higher resolution.
  %%When droplets are counted in these low-resolution simulations, an exceedingly large number of droplets is found, precluding a meaningful analysis. }
  \red{When droplets are counted in these low-resolution simulations, an exceedingly large number of small droplets resulting from numerically dominated breakup, -- so called debris, flotsam or jetsam or wisp droplets --  is found, precluding a meaningful analysis in the small droplet range. This type of debris droplet is absent or rare in pure Level Set simulations.}

%% Consequently, high-shear regions are more numerous and difficult to capture than in case 1. The vorticity in fluid 2 is modulated by the presence of the light droplets of fluid 1 that first follow the motion of fluid 2 at short times, owing to inertia and viscous effects, and then rise under buoyancy effects. The finer the grid, the smaller the droplets captured by the computation. This feature has a direct influence on the intensity and density of vortical regions in the flow, since the latter are closely related to the phase distribution, \textit{i.e.} to the dispersion of fluid 1 in the present case. Therefore the behavior of the volume-averaged enstrophy depicted above is a direct consequence of this grid-dependent population of droplets.

\section{Summary and future work}

A computational benchmark based on the phase inversion of two immiscible fluids of different densities confined in a closed cubic box has been set up. The purpose of this benchmark was to check the capabilities and limitations of current codes and grid resolutions. %with the ultimate goal of producing numerical data of a quality similar to those provided by a DNS of single-phase turbulence.
%Several numerical approaches  based on the one-fluid model and routinely used to compute two-phase incompressible flows have been used and compared. They include the Volume-Of-Fluid and Level-Set approaches for interface tracking, the Ghost-Fluid technique for the capture of interfacial jumps of mass and momentum, as well as several variants of Navier-Stokes solvers such as the time-splitting and augmented-Lagrangian algorithms. All these approaches and methods are \textit{a priori} suitable for simulating the various configurations encountered in phase-inversion problems on an arbitrary grid.
{The first observation made with the reference code, DyJeAT, is that the total mechanical energy converges easily with grid resolution and can be rescaled so that it exhibits a very similar behavior in both cases. This effect was unexpected and it would be interesting to see if it applies to a wider range of cases.
The second observation, put simply, is that the various codes, all run at much lower resolution than the reference, reproduce easily the results of the reference for quantities tied to the large scales, such as kinetic and potential energies, but fail to reproduce quantities related to the small scales, such as the PDF of droplet sizes or the enstrophy integral. } 

 The first problem we considered, case 1, is characterized by an integral scale Reynolds number Re = \Refi. Computational results have been analyzed by considering the time evolution of several volume-averaged indicators, namely the potential and kinetic energies in each fluid, the relative volume of light fluid in the top part of the box and the enstrophy in each fluid. The results reveal that all codes provide close evolutions for the first three quantities, indicating that reliable predictions are obtained when quantities essentially determined by large-scale motions are considered. The long-term decay rates of the kinetic energies have been found to be in good agreement with the viscous Stokes law (in fluid 1) and the turbulent decay law (in fluid 2) discussed in Appendix \ref{AppendixA}, respectively. In addition, all codes agree on the sloshing frequency of the light fluid. The situation has been found to be much more problematic when enstrophy is considered: all codes provide markedly different values of the enstrophy maximum (although they agree on the time at which this maximum occurs) and none of them converges when the grid is refined.

%The analysis revealed that this failure is due to the use of too coarse grids that are unable to capture properly the vortical layers induced by the interface motion, especially in the vicinity of the walls.\hskip 7pt
A second phase inversion test case, involving a higher Reynolds number, \Re = \Refii \mbox{}, was subsequently considered. The same general conclusions apply to this configuration. All codes provide converging predictions as long as only kinetic and potential energies are considered, whereas large discrepancies are observed on enstrophy. Here again, the true DNS conditions are not satisfied and much finer grids should be considered to expect convergence on quantities such as enstrophy and dissipation which are governed by small-scale processes, especially multiple break-up events. It has to be noticed that despite the discrepancies observed on enstrophy, approximate convergence is obtained for PDFs of droplet sizes as soon as a $1024^3$ grid is used. \textcolor{black}{Clearly, case 2 is an implicit LES rather than a DNS. As a conclusion, this benchmark {illustrates} the fact that implicit LES simulations on fine grids allow to reach convergence on first-order moments such as kinetic or potential energies while they fail to converge as soon as small-scale statistics are investigated. Explicit LES models may be implemented in this case \cite{ Wojtek, Labourasse, Vincent64, Vincent4, VincentLES, MahdiLES}, or much finer meshes should be employed to achieve a true DNS.}

Future work of some of the partners will first be devoted to the consideration of cases with smaller $\Re$ and $\We$ in order to perform true DNS with convergence of the small scales.  Multiscale Eulerian VOF or Level Set approaches coupled to a Lagrangian description of small droplets  may also be introduced by some of them with the goal of reaching convergence on all physical quantities on ``reasonable'' grids. Last, adaptive mesh refinement (AMR) techniques will be considered to concentrate numerical efforts in flow regions where the vorticity magnitude is expected to approach its maxima.

\section{Acknowledgements}

This work was part of the MODEMI ANR project (\textit{Mod\'elisation et Simulation Multi-\'echelle des Interfaces}, ANR-11-MONU-0011) devoted to the multiscale modeling of two-phase flows. We thus acknowledge the support of the Agence Nationale de la Recherche (ANR) for the I2M, D'ALEMBERT, CORIA and IMFT laboratories concerning the Multiscale Modeling of Interfaces.
The authors wish to thank the Midi-Pyr\'en\'ees and Aquitaine
Regional Councils for the financial support dedicated to a PhD thesis
at ONERA and IMFT and a 256-processor cluster investment, located in
the TREFLE team of the I2M laboratory. The D'ALEMBERT team benefited from the ERC grant TRUFLOW. We are grateful for access to the
computational facilities of the French CINES
(National computing center for higher education) and \red{TGCC, notably Irene-Rome, granted by GENCI} under project numbers x20142b6115, x20152b6115, x20162b6115, A0012b06115 \red{and A0092B07760}.
The project also benefited from the PRACE grant TRUFLOW \red{number 2020225418} for a large amount of CPU  hours \red{on TGCC Irene-Rome} in 2021.
This work was also partly granted access to the HPC resources of CALMIP under the allocation 2013-P0633.  The CORIA team would like to express its gratitude to the CRIANN
(Centre R\'egional Informatique et d'Applications Num\'eriques de Normandie, www.criann.fr)
for providing CPU resources. \red{We thank the technical and administrative teams of these supercomputer centers and agencies for their kind and efficient help. }

\appendix %This fixes the numeration of appendices (WA)
\section{Appendix A: Scaling laws for the kinetic energy decay}\label{AppendixA}

After the acceleration stage induced by buoyancy forces, during which most of the lighter fluid goes to the top part of the box, the phase inversion problem is characterized in a second stage by a wavy behavior which makes it look quite similar to a sloshing flow progressively damped by shearing and viscous effects. The classical analysis of Stokes regarding the viscous damping of gravity waves (see \textit{e.g.} \cite{Lamb}, pp. 623-624), makes it possible to predict the time evolution of the kinetic energy in a weakly viscous flow driven by a surface wave. First, the time evolution of the mechanical energy, which is the sum of the potential and kinetic energies, is known to result from the internal viscous dissipation, so that
\begin{equation}
\frac{{\rm d} E_m}{{\rm d}t}=-\frac{1}{2} \mu \int_\Omega \left(\nabla {\bf u} + (\nabla {\bf u})^\text{T} \right)^2 \,{\rm d}V \,.
\end{equation}
If we assume the shear layer at the free surface to have negligible effect owing to the moderate velocity gradients expected in this region, the flow can be considered irrotational. Then the velocity potential, $\Phi$, and rate of change of the mechanical energy, $E_m$, read
\begin{eqnarray}
\Phi = \Phi_0 e^{kz}cos(k x - \omega t ) \,, \label{eqA0}\\
%\frac{\partial u_i}{\partial x_k} = \frac{\partial u_k}{\partial x_i} = \frac{\partial^2 \phi}{\partial x_i \partial x_k} \label{eqA00}\\
\frac{{\rm d} \overline{E_m}}{{\rm d}t}=-8\mu k^4 \int_\Omega \overline{\Phi^2} \,{\rm d}V \,,\label{eqA1}
\end{eqnarray}
where $k$ and $\omega$ are the wave number and radian frequency, respectively, and the overbar denotes the time average value. In the framework of linear wave theory, the potential and kinetic energies are equal. However the wave amplitude is not small in the present phase inversion problem. Nevertheless, as soon as most light fluid stands in the top part of the box, the potential energy stays almost constant, making the time variation of $E_m$ mainly governed by that of the kinetic energy. Therefore, still in the linear approximation, we can approximately write
\begin{equation}
\overline{E_m} = \frac{1}{2} \rho \int_\Omega \overline{{\bf u}^2}\, {\rm d}V\approx \frac{1}{2} \rho \int_\Omega {\overline{\nabla \Phi\cdot\nabla\Phi}} \,\,{\rm d}V = \rho k^2 \int_\Omega \overline{\Phi^2} \,{\rm d}V \label{eqA2} \,.
\end{equation}

Now, two markedly different flow situations can be met in the phase inversion problem, namely a ``gentle" configuration in which break-up and coalescence events of the light fluid scarcely occur and a ``violent" configuration in which such events are much more numerous. Combining (\ref{eqA1}) and (\ref{eqA2}), the time evolution of the kinetic energy in the first regime is found to obey
\begin{equation}
\overline{E_m} = \frac{1}{2} \rho \int_\Omega\overline{ {\bf u}^2} \,{\rm d}V = K e^{-8\nu k^2 t} = K e^{-8\nu \frac{\omega^4}{g^2} t}\,,
\label{expo}
\end{equation}
where $K$ is a constant, $\nu=\mu/\rho$, and use has been made of the dispersion relation $\omega^2 =gk$.  \\
Let us now consider the ``violent" configuration. In this case, averaging throughout the whole volume $\Omega$, (\ref{eqA0})-(\ref{eqA1}) simply yields
\begin{eqnarray}
\widetilde{{\bf u}^2}=\widetilde{\nabla\Phi\cdot\nabla\Phi}=k^2 \widetilde{\Phi^2}\,, \\
\frac{{\rm d} \overline{E_m}}{{\rm d}t}=-8\mu k^2 \widetilde{{\bf u}^2}\,,
\end{eqnarray}
where $\widetilde{\cdot}$ denotes the mean value over $\Omega$. In other words, one has
\begin{equation}
\frac{{\rm d} \widetilde{{\bf u}^2}}{{\rm d}t}=-16 \nu k^2 \widetilde{{\bf u}^2}\,. \label{eqA3}
\end{equation}
In this configuration, the flow is expected to be turbulent and the molecular viscosity is no longer relevant to estimate the damping rate of the flow.  As a crude surrogate, we can introduce an effective turbulent viscosity $\nu_t$ scaling as $l \widetilde{\bf u}$, where $l$ is a characteristic length of the large-scale flow, which can be taken as the box size. Replacing $\nu$ by $\nu_t$ in (\ref{eqA3}), we then obtain
\begin{equation}
\frac{{\rm d} \widetilde{{\bf u}^2}}{{\rm d}t}=-16 l k^2 \widetilde{{\bf u}^3} \,.
\end{equation}
Assuming the volume-averaged velocity to follow a power law, \textit{i.e.} $\widetilde{{\bf u}}$ to be of the form $\widetilde{{\bf u}_0} t^{-n}$, yields $n=1$, from which we infer
%\begin{eqnarray}
%\frac{{\rm d} t^{-2n}}{{\rm d}t}=-16 l \alpha k^2 t^{-3n} \\
%n=1
%\end{eqnarray}
\begin{equation}
\overline{E_m}\sim t^{-2} \,.
 \label{LanenLaw}
\end{equation}
This prediction is to be compared with the exponential decay predicted by (\ref{expo}) in the ``gentle" regime. The result (\ref{LanenLaw}) is reminiscent of the approximate decay law of the kinetic energy in decaying homogeneous isotropic turbulence (\textit{e.g.} \cite{Sagaut2}). This is expected since, according to Taylor's estimate, the dissipation rate $\epsilon$ is known to scale as $u_0^3/l_0$, where $u_0$ and $l_0$ stand for the large-scale velocity and length scales, respectively. Therefore, when $l_0$ is constant, (\ref{LanenLaw}) is immediately recovered and the dissipation rate is predicted to decay as $t^{-3}$.
%\begin{equation}
%\overline{E_m}\sim e^{-8\mu \frac{\omega^4}{g^2} t} \label{LanLaw}
%\end{equation}

\section{Appendix B: influence of viscosity on the flow dynamics}

\label{AppendixB}

In order to better understand the origin of code- and grid-dependencies
of the volume-averaged enstrophy in cases 1 and 2, we performed an extra
series of computations with DyJeAT with the same viscosity in both
fluids. In this way, any possible influence of the averaging procedure
selected to compute the local viscosity as a function of the volume
fraction and of the numerical treatment of the viscosity jump in the
interfacial grid cells is removed. We make use of the physical
parameters of case 1, except for viscosity which is set to $0.1 Pa.s$
(case 1a), $0.01 Pa.s$ (case 1b) and $0.001 Pa.s$ (case 1c),
respectively. %The simulations are only carried out with the DyJeAT code, as it allows us to use the largest grids at a reasonable cost. 
The convergence study was performed on three different grids,
namely $128^3$, $256^3$ and $512^3$ in case 1a, whereas four grids
ranging from $128^3$ to $1024^3$ were considered in cases 1b and 1c.\\

\begin{figure}[ht!]
\begin{center}
  \includegraphics[width=6cm]{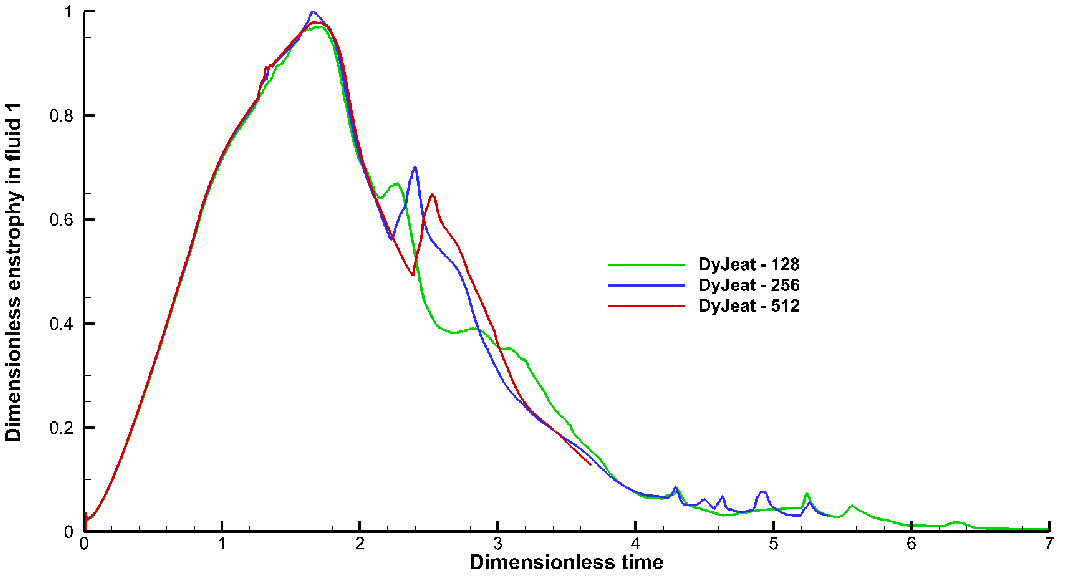}
  \includegraphics[width=6cm]{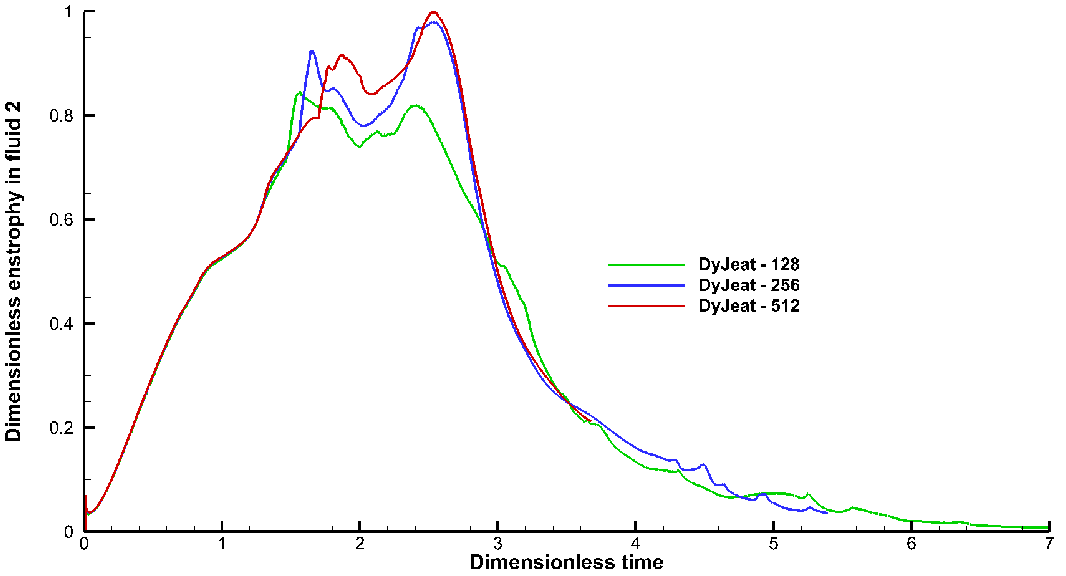}
  \includegraphics[width=6cm]{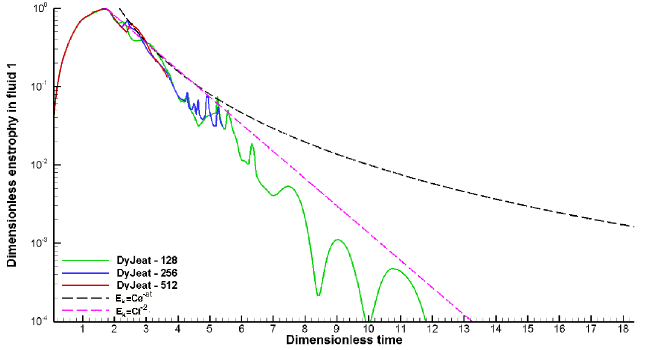}
  \includegraphics[width=6cm]{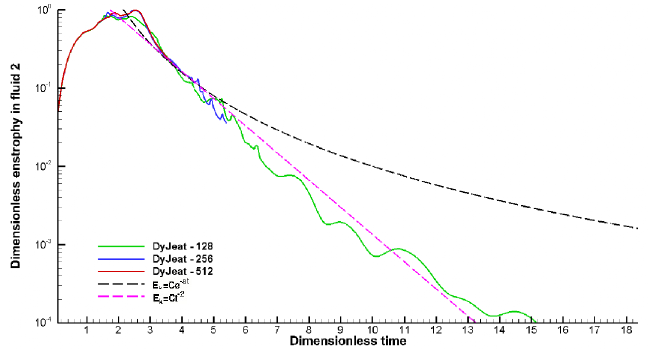}
  \end{center}
\caption {Grid convergence of enstrophy for case 1a in fluid 1 (left) and fluid 2 (right) - linear-linear (top) and log-linear (bottom) coordinates.}
\label{fig4.15}
\end{figure}

\begin{figure}[ht!]
\begin{center}
  \includegraphics[width=6cm]{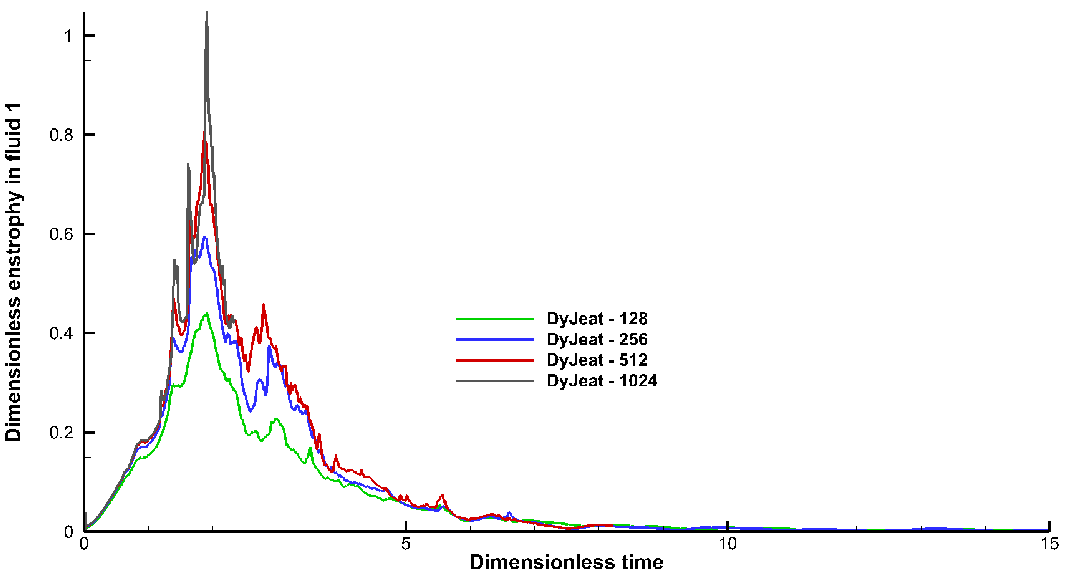}
  \includegraphics[width=6cm]{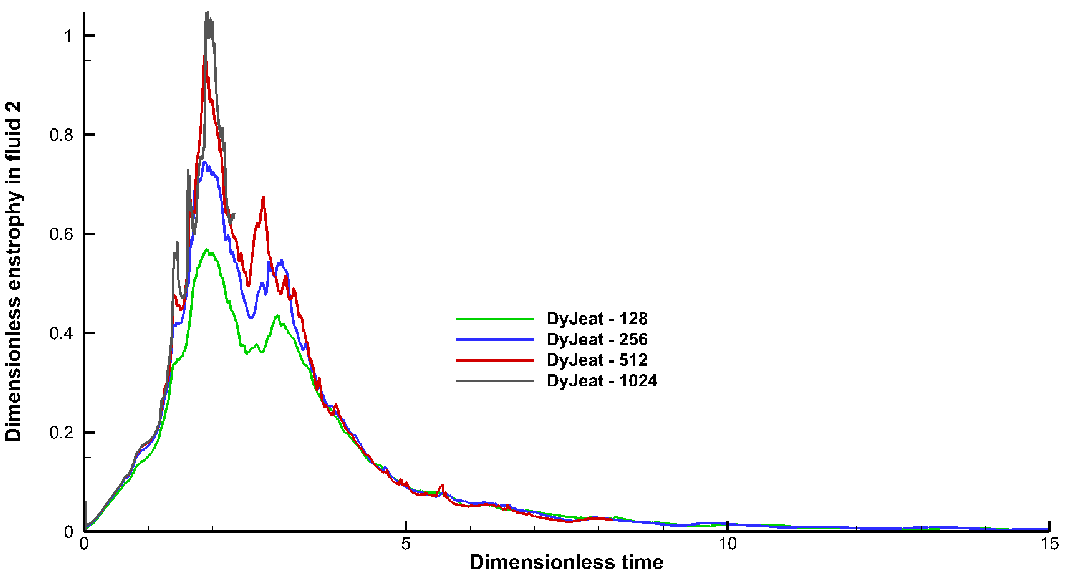}
  \includegraphics[width=6cm]{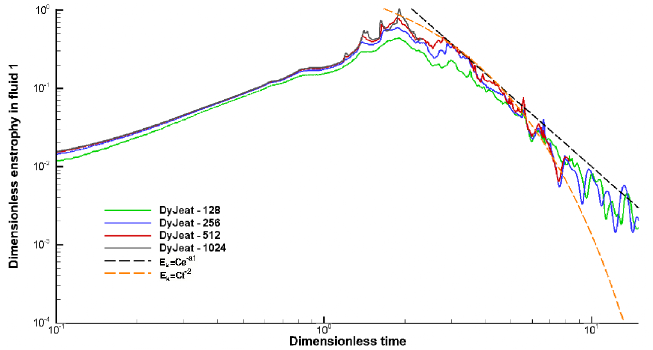}
  \includegraphics[width=6cm]{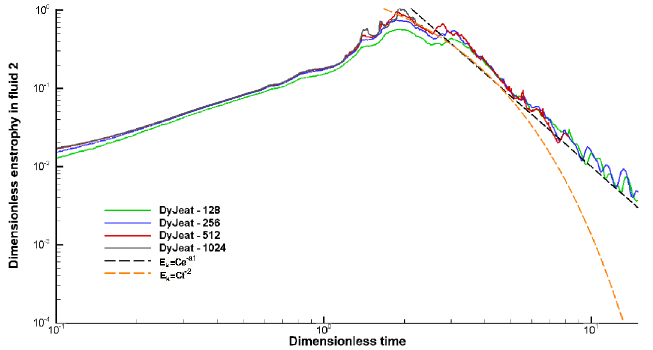}
  \end{center}
\caption {Grid convergence of enstrophy for case 1b in fluid 1 (left) and fluid 2 (right) - linear-linear (top) and log-log (bottom) coordinates.}
\label{fig4.16}
\end{figure}

\begin{figure}[ht!]
\begin{center}
   \includegraphics[width=6cm]{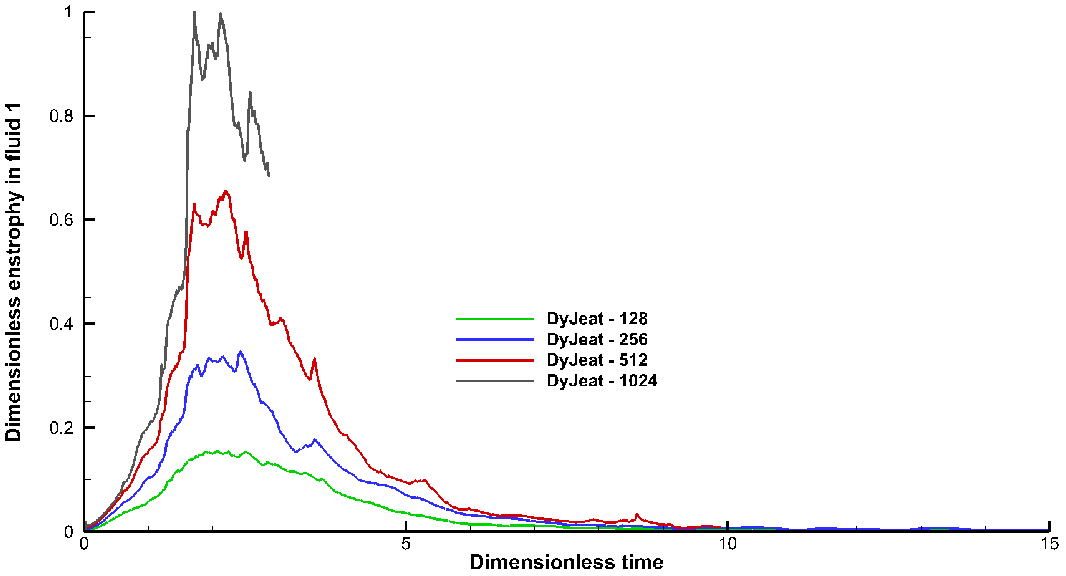}
   \includegraphics[width=6cm]{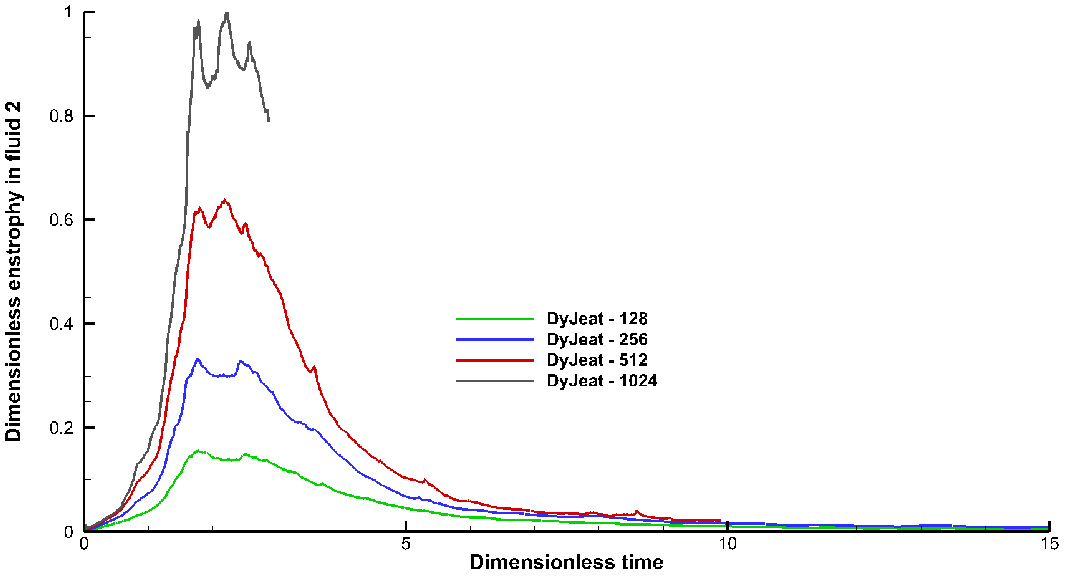}
   \includegraphics[width=6cm]{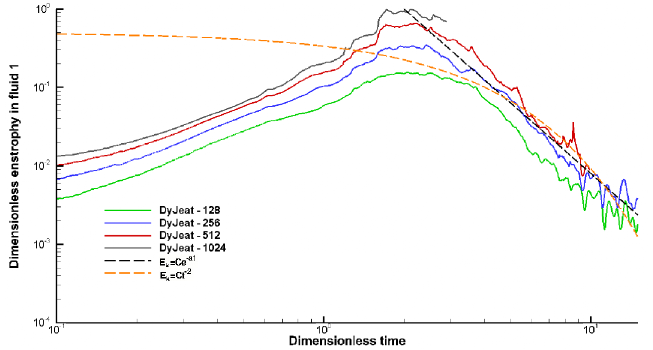}
   \includegraphics[width=6cm]{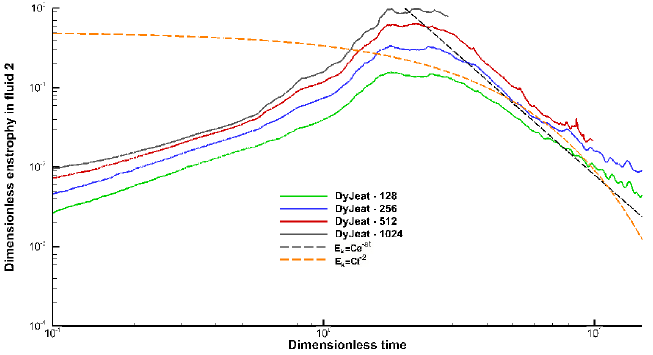}
  \end{center}
\caption {Grid convergence of enstrophy for case 1c in fluid 1 (left) and fluid 2 (right) - linear-linear (top) and log-log (bottom) coordinates.}
\label{fig4.17}
\end{figure}

The evolution of the volume-averaged enstrophy is reported in Figures
\ref{fig4.15}, \ref{fig4.16} and \ref{fig4.17} for cases 1a, 1b and
1c, respectively. In case 1a, the Reynolds \textcolor{black}{number is $130$
and the enstrophy while still not converging  as the grid is further refined,
has less violent excursions}. In this
configuration, the enstrophy decay at large time follows an
exponential law, a behavior typical of the viscous decay of gravity
waves (see Appendix \ref{AppendixA}). In case 1b, grid convergence is
again \textcolor{black}{not} achieved in both fluids, the Reynolds number in fluid 2 being
$1300$. At large time, the enstrophy obeys a $t^{*-3}$ decay law
typical of turbulent conditions (see Appendix \ref{AppendixA}). In
case 1c, enstrophy convergence is not achieved in either fluid.
The Reynolds number is {\Refi} and it is observed that,
even though the results concerning energy and relative volume in the
upper part of the box converge, as shown with case 1, the grid is not
thin enough to capture the small-scale eddies. Again, the decay law at
large times is found to correspond to turbulent conditions.\\

\bibliography{p} 

\end{document}